\documentclass[conference,compsoc]{IEEEtran}
\usepackage{times}
\usepackage[utf8]{inputenc}
\usepackage{amssymb}
\usepackage{amsmath}
\usepackage{amsthm}
\usepackage{booktabs}
\usepackage{graphicx}
\usepackage[usenames, dvipsnames]{xcolor}
\usepackage{multirow}
\usepackage{tikz}
\usepackage{colortbl}
\usepackage{multicol}
\usepackage{titlesec}
\usepackage[numbers,sort]{natbib}

\titlespacing\section{0pt}{7pt plus 4pt minus 2pt}{5pt plus 2pt minus 2pt}
\titlespacing\subsection{0pt}{7pt plus 4pt minus 2pt}{6pt plus 2pt minus 2pt}
\titlespacing\subsubsection{0pt}{7pt plus 4pt minus 2pt}{5pt plus 2pt minus 2pt}

\makeatletter
\newcommand\footnoteref[1]{\protected@xdef\@thefnmark{\ref{#1}}\@footnotemark}
\makeatother

\newcommand{\shawn}[1]{{\color{black} #1}}

\newcommand{\para}[1]{{\vspace{4pt} \bf \noindent #1 \hspace{4pt}}}

\newcommand{\htedit}[1]{{\color{black} #1}}

\newcommand{\erev}[1]{{\color{black} #1}}

\newcommand{\rev}[1]{{\color{black} #1}}

\DeclareRobustCommand{\crc}[1]{
  \mathrel{
    \tikz[baseline=(char.base)]{
      \node[shape=circle,draw,inner sep=0.8pt] (char) {#1};
    }
  }
}

\DeclareRobustCommand{\crcs}[1]{
  \mathrel{
    \tikz[baseline=(char.base)]{
      \node[shape=circle,draw,inner sep=0.2pt] (char) {#1};
    }
  }
}

\usepackage{relsize}

\newcommand*\fcirc[1][1ex]{\tikz\fill (0,0) circle (#1);}
\newcommand{\ecirc}[1][1]{\text{\larger[#1]{\textbf{?}}}}
\newcommand*\gcirc[1][1ex]{\tikz\fill[gray!50] (0,0) circle (#1);} 

\newcommand{\eg}{{e.g.,\ }}
\newcommand{\ie}{{i.e.\ }}
\newcommand{\etal}{{\em et al.\ }}

\newcommand{\pr}{$P$}
\newcommand{\adv}{$\mathbb{F}$}

\newenvironment{packed_itemize}{
\begin{list}{\labelitemi}{\leftmargin=1.em}
\setlength{\itemsep}{3pt}                                                           
\setlength{\parskip}{0pt}
\setlength{\parsep}{0pt}
\setlength{\headsep}{1pt}
\setlength{\topskip}{0pt}
\setlength{\topmargin}{0pt}
\setlength{\topsep}{0pt}
\setlength{\partopsep}{0pt}
}{\end{list}}                                                                                                                                                 
\newenvironment{packed_enumerate}{  
\begin{enumerate}
  \setlength{\itemsep}{3pt}
  \setlength{\parskip}{0pt}
  \setlength{\parsep}{0pt}
  \setlength{\headsep}{0pt}
  \setlength{\topskip}{0pt}
  \setlength{\topmargin}{0pt}
  \setlength{\topsep}{0pt}
  \setlength{\partopsep}{0pt}
}{\end{enumerate}} 

\title{SoK: Anti-Facial Recognition Technology}
\vspace{-2cm}
\author{\IEEEauthorblockN{Emily Wenger}
\IEEEauthorblockA{\textit{University of Chicago}\\
ewenger@uchicago.edu}
\and
\IEEEauthorblockN{Shawn Shan}
\IEEEauthorblockA{
\textit{University of Chicago}\\
shawnshan@cs.uchicago.edu}
\and
\IEEEauthorblockN{Haitao Zheng}
\IEEEauthorblockA{
\textit{University of Chicago}\\
htzheng@cs.uchicago.edu}
\and
\IEEEauthorblockN{Ben Y. Zhao}
\IEEEauthorblockA{
\textit{University of Chicago}\\
ravenben@cs.uchicago.edu}
}
\begin{document}

\maketitle

\begin{abstract}
   The rapid adoption of facial recognition (FR) technology by both
   government and commercial entities in recent years has raised concerns
   about civil liberties and privacy. In response, a broad suite of so-called
   ``anti-facial recognition'' (AFR) tools has been developed
   to help users avoid unwanted facial recognition. The set of AFR tools proposed in the
   last few years is wide-ranging and rapidly evolving, necessitating a step back to
   consider the broader design space of AFR systems and long-term challenges.
   This paper aims to fill that gap and provides the first
   comprehensive analysis of the AFR research landscape. Using the
   operational stages of FR systems as a starting point, we create a
   systematic framework for analyzing the benefits and tradeoffs of
   different AFR approaches. We then consider both technical and social challenges
   facing AFR tools and propose directions for future research in this
   field. 
   
  \end{abstract}

\section{\bf Introduction}
\label{sec:intro}

In recent years, facial recognition systems have accelerated their growth in
scale and reach, and are becoming an increasingly ubiquitous part of our daily lives.
As a result, the majority of citizens in the world's most populous countries are
already enrolled in one or more
facial recognition systems, whether they know it or not.  For example,
in the United States, nearly 200
million residents are already enrolled in the FBI's facial recognition
database, which was built by leveraging FBI's access to driver license photos
in many 
states~\cite{perpetual}. In China, a well-known surveillance system uses
facial recognition to monitor civilian behavior and enforce the social credit
score system~\cite{uighur_surveill, mozur2019one}. In Russia, authorities acquired 100,000+ cameras in
Moscow to build a facial recognition-based COVID quarantine enforcement
system~\cite{abc_russia}. Beyond government use cases, facial recognition systems are now regularly used for myriad purposes, including authenticating
 travelers at airports and employees entering corporate offices.

The advancements that paved the way to real-world facial recognition systems have also opened the door to their potential misuse and
abuse. With moderate resources, an individual or institution, public or
private, can now extract training data from social media and online sources to
build facial recognition models capable of recognizing large groups of users. In 2020, New York
Times journalist Kashmir Hill confirmed the potential for facial recognition misuse
when she profiled Clearview.AI, a private for-profit company that
scraped over 3 billion images from ``public sources''  to build a facial
recognition system that recognized hundreds of millions of private
citizens~\cite{hill_clearview}, without their knowledge or
consent. Clearview and companies like it could enable surveillance and
tracking by anyone willing to pay\footnote{Multiple countries are pursuing
  inquiries into Clearview's business model, and Canada has already denounced
  it as ``illegal''~\cite{canada_nyt}.}.  %In addition to leveraging images shared online, 
  Other reports have detailed how photos taken in unexpected places -- airports, city streets, government buildings,
schools, corporate offices -- end up in facial recognition systems
without subjects' knowledge or consent~\cite{perpetual, biometric_reg,
  nist_fvr_test, facesurveillance, china_realestate_fr, india_monitoring}.

Despite backlash against intrusive facial recognition
systems~\cite{aws_banned_police, ibm_ceo_letter, banthescan, bigbrother},
there are few commercial or legislative  tools available to protect users against them. While big
tech has begun to self-regulate~\cite{meta_fr} and openly
\htedit{called} for legislation (\eg~\cite{aws_banned_police,ibm_ceo_letter}), legislative efforts to
regulate facial recognition remain scarce.  In their place, a
cottage industry of anti-facial recognition (AFR) tools has emerged. 
AFR tools are designed to target different parts of facial recognition
systems, from data collection,  model training to run-time inference, with the
unified goal of preventing successful recognition by unwanted or unauthorized models.

\begin{figure*}[t]
\begin{center}
    \centering \includegraphics[width=0.8\textwidth]{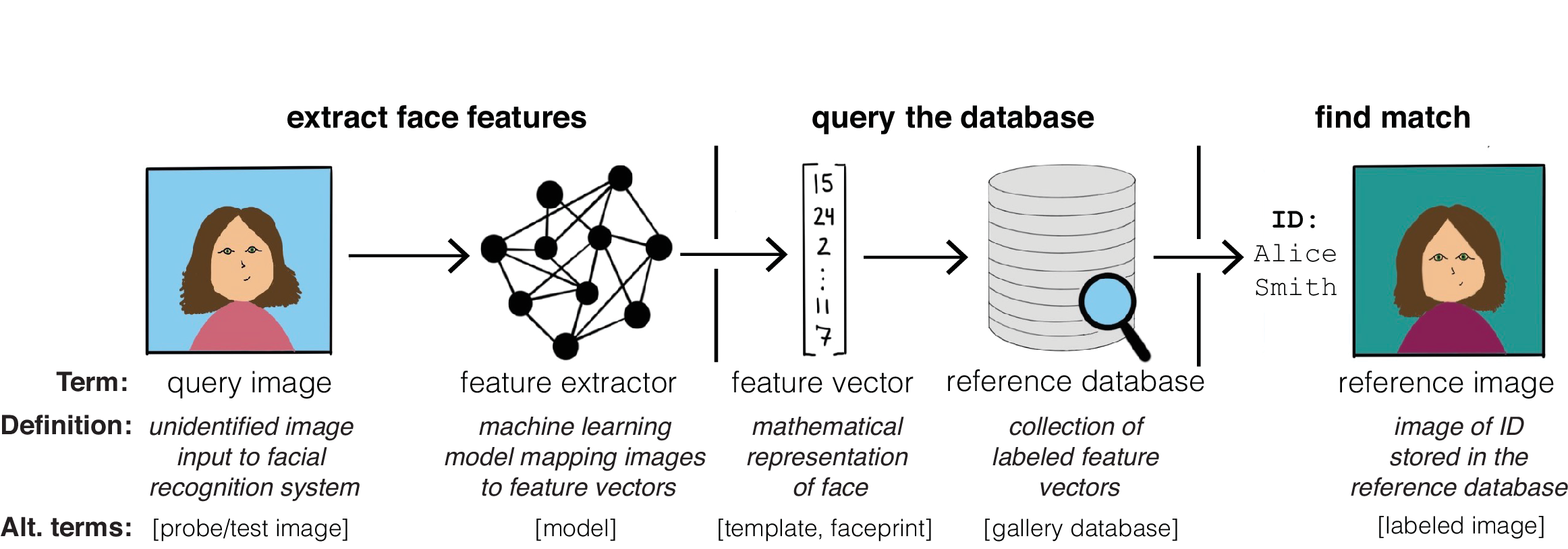}
 \vspace{-0.05in} \caption{The workflow of how facial recognition
      systems recognize a human face in an input image, along with the
      corresponding terminology. (a): A query image, after being submitted to the system, is passed to the feature extractor to
      produce a feature vector;  (b): this feature vector is used to query a reference database
    of labeled feature vectors;  (c):  if the query feature vector matches
    a labeled feature vector in the database, the label
    is used to find a reference image, and the
    system outputs the reference image and the identity (i.e. Alice Smith
    in this example). 
  }
    \label{fig:overview} \vspace{-0.2in}
\end{center}
\end{figure*}

\looseness=-1 AFR tools have also attracted significant attention from the
  research community. In the last 12 months, more than a dozen AFR tools have
been proposed (e.g.,~\cite{deepblur, identitydp, fawkes, unlearnable, foggy, lowkey,
  preventing, wu2020making, treu2021fashion, xu2020adversarial, faceoff,
  xue2020socialguard, adversariallight, advhat, cameraadversaria,
  vincent2021can}).   While most are constrained to research prototypes, a few
of these tools have produced public software releases and gained significant
media attention~\cite{fawkes, lowkey,cvdazzle}.

\looseness=-1 Proposals in the rapidly growing collection of AFR tools differ widely
in their assumptions and techniques, and target different pieces of the facial recognition pipeline. There is a need to better understand their
commonalities, to highlight performance tradeoffs, and to identify unexplored
areas for future development. Existing
  surveys~\cite{meden2021privacy,padilla2015visual} on facial privacy
  issues do not \rev{consider user-centric AFR tools. They instead discuss privacy-preserving techniques that surveillance system operators could employ, a related but separate line of work to that addressed here (see \S\ref{sec:discussion})}. %\cite{meden2021privacy} summarizes proposals that modify

In this paper, we address this need by providing \htedit{a common framework for analyzing a wide range of AFR systems.}
More specifically, we make the following contributions:

\begin{packed_itemize}
\item {\bf Taxonomy of targets in facial recognition systems:} AFR systems target a wide range of components in the facial recognition process. Using
  a generalized version of the FR data pipeline, we provide a framework
  for reasoning broadly about existing and future AFR work.
\item {\bf Categorization and analysis of AFR tools:} We take the current body of work
  on AFR tools, and categorize and analyze them using our proposed
  framework.
\item {\bf Mapping design space based on desired properties:} We identify a
  core set of key properties that future AFR systems might optimize for in
  their design, and provide a design roadmap by discussing how and if such
  properties can be achieved by AFR systems that target each stage in our
  design framework. 
\item {\bf Open challenges:} We use our framework to identify
  significant challenges facing current AFR systems, as well as 
  directions for potential solutions. %\vspace{3pt}
\end{packed_itemize}

The rest of the paper proceeds as follows. We begin by providing operational details of
real-world facial recognition systems (\S\ref{sec:back}), including real-world deployment scenarios
and key technical components. We then present the \htedit{motivation
  and threat model} of AFR tools (\S\ref{sec:threat}), and our framework for analyzing existing AFR tools (\S\ref{sec:stages}). We
then discuss existing AFR proposals
targeting each stage, i.e., {\em data collection} (\S\ref{sec:collection}), {\em
  data processing} (\S\ref{sec:preprocess}), {\em feature extractor
training} (\S\ref{sec:poisongal}), {\em identity creation}
(\S\ref{sec:poison}), and {\em query matching} (\S\ref{sec:evade}).
\htedit{Finally, we identify desirable properties of effective AFR systems,
  and map them to points in the design space
  (\S\ref{sec:tradeoffs}). Finally, we discuss challenges and
  directions for AFR research (\S\ref{sec:future}).}

\looseness=-1 \para{\em Unresolved Ethical Questions:} The broad deployment of
facial recognition systems (and by extension, AFR systems) is fraught with
ethical challenges, not the least of which are significant
biases against women and people of color~\cite{gendershades}. While we
discuss ethical tensions surrounding AFR systems in~\S\ref{subsec:social}, we
do not make assertions about how (and whether) AFR tools should be
used. Development and adoption of AFR tools are driven by backlash against
biased and misused facial recognition systems. Though their legal and ethical implications are yet-unknown, we believe that AFR
tools are here to stay. Consequently, 
% This paper is written in light of the reality: AFR tools exist, even as we
an analysis of their strengths and limitations is crucial to
advancing the ongoing debate about their use and the place of
facial recognition in our world.

\vspace{-0.1cm}
\section{{\bf Facial Recognition: Terminology, Design Stages and Deployment}}
\label{sec:back}

To provide context for later discussions, we now give a high-level
overview of 
today's facial recognition (FR) systems and their real-world
implementations.  Our goal is to describe modern FR 
systems targeted by today's AFR systems,  their key operational
stages, and
how these FR systems are being deployed around the globe.  Together,
these provide a framework that we will use for analyzing
AFR systems later in \S\ref{sec:stages},  by examining critical points of direct interaction between users and FR
systems.  

\vspace{-0.2cm}
\subsection{{\bf Modern FR Systems}}
\vspace{-0.1cm}

FR systems identify people by their facial characteristics, generally by comparing an unknown face in an image (or a video) against a database of known faces. The technology has evolved significantly over the past two decades, resulting in many design
variants~\cite{facerecreview}.  Today, the state-of-the-art and widely adopted FR systems employ deep neural networks (DNNs) to extract unique features from a given face.  Since existing AFR
systems mainly target these modern FR systems\rev{, this work focuses on DNN-based FR/AFR systems. The main differences between older and newer FR methods lie in (1) the feature extraction methods (e.g. statistical methods like PCA or LDA~\cite{turk1991eigenfaces, belhumeur1997eigenfaces} vs. DNN-based feature extractors) and (2) scale However, the fundamental FR stages remain the same in both older and newer FR systems – face images must still be collected, processed, and recognized. Consequently, the framework laid out in this paper could be applied to older FR/AFR systems if desired.}

\begin{figure*}[t]
  \centering
  \includegraphics[width=0.95\textwidth]{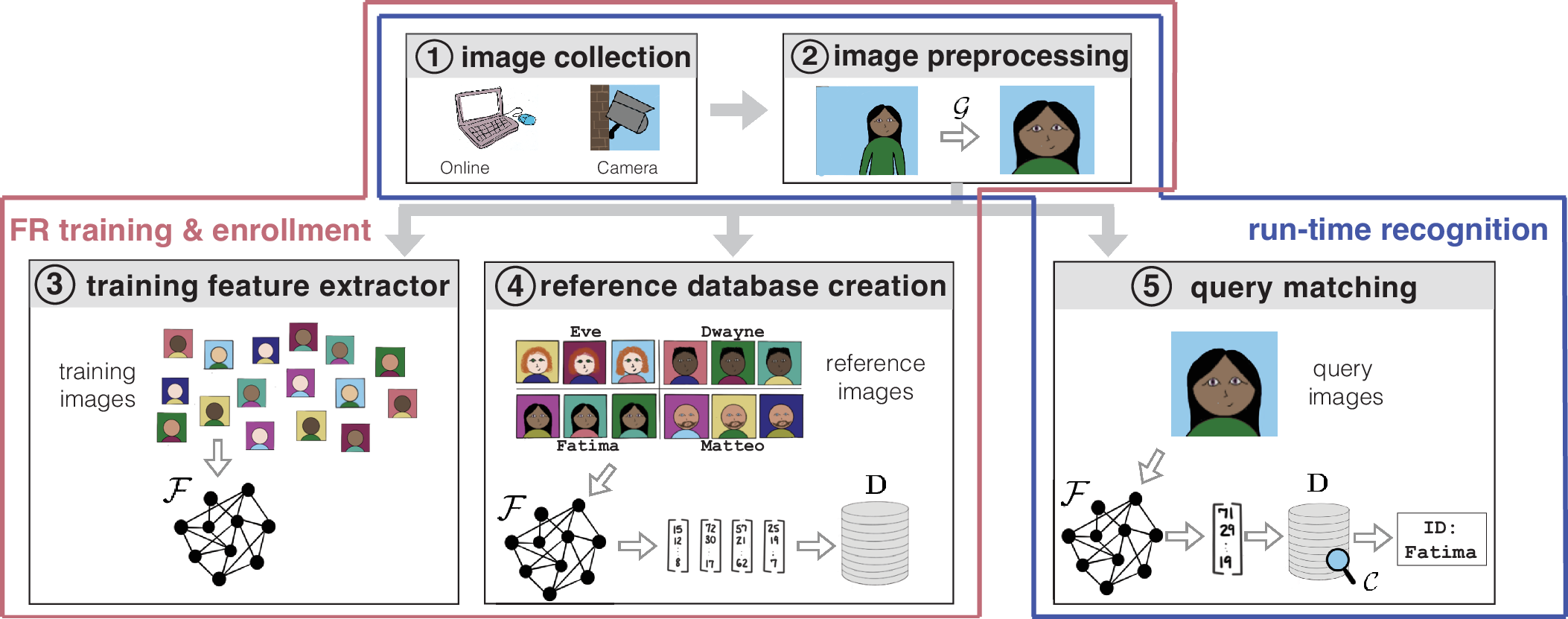}
  \caption{We propose to divide the operational pipeline of a FR system \adv$=\{\mathcal{G}, \mathcal{F}, \mathcal{C}, \mathbf{D}\}$
  into a set of five operational stages $\crc{1}$ to $\crc{5}$. They
  encompass the five critical points of direction interaction between
  users and FR systems. Later we will use this framework for
  analyzing AFR systems.}
\label{fig:dataflow}
\end{figure*}

\para{Terminology.} In this paper, we represent a modern FR system as  \adv~=
$\{\mathcal{G}, \mathcal{F},  \mathcal{C}, \mathbf{D}\}$, whose goal is
to associate a query image $x_I$ with its true identity
$I$.  Specifically,

\begin{packed_itemize}

\item \textit{Query image} ($x_I$): a face image to be
  identified by \adv.

\item \textit{Preprocessing engine} ($\mathcal{G}$):   a processing function that prepares raw face images for the FR task,
  e.g., detecting and cropping out individual faces.

\item \textit{Face feature extractor} ($\mathcal{F}$): a DNN that
  converts a face image into a \textit{feature vector}, a mathematical representation of
  the person's unique facial features.

 \item \textit{Reference database} ($\mathbf{D}$):   a collection of face images and their feature vectors of known
identities, e.g., $x^R_I$ \rev{(ground truth images of user $I$)} and $\mathcal{F}(\mathcal{G} (x^R_I)) = v_I$. %are stored for user $I$. 

 \item \textit{Run-time face classifier} ($\mathcal{C}$): this
   function runs a 
   query search to match the query image $\mathcal{F}(\mathcal{G}
   (x_I))$ against $\mathbf{D}$. If the closest feature vector $v_I$ is sufficiently similar, then the query image is
   identified as $I$. Ideally, it should produce 
   \rev{$\mathcal{C}(\mathcal{F}(\mathcal{G}
   (x_I)), \mathbf{D})=\mathcal{C}(\mathcal{F}(\mathcal{G} (x^R_I)), \mathbf{D})$, where $x_I$ is a previously unidentified image of $I$ and $x^R_I$ is a ground-truth reference image of $I$}. 

 \end{packed_itemize}
 %\vspace{-0.3cm}
\looseness=-1 It should be noted that the terminology used to describe a FR system can vary
across the literature. We list some alternative terms  in
Figure~\ref{fig:overview}. The terms we choose to use in this paper are,
we believe, most familiar to the security community.
%, which is the primary audience for this paper. 

\para{Face recognition vs. face verification.}
Here, we note the distinction 
between {\em face recognition} and {\em face verification}. Face
verification is widely used to authenticate
users on mobile devices (\eg FaceID on iPhones) by comparing a user's
face feature to the stored face feature of the 
authorized user.  While the two systems apply similar techniques to analyze face
images, facial verification systems require \textit{user consent} for
deployment while many FR systems operate without user consent.  As such, most AFR
systems target FR systems rather than face verification systems. {\em
  Furthermore, face verification systems can be viewed as a special
  case of FR systems, where the reference database only has a single
  user.} Therefore, in this paper we do not explicitly consider
facial verification or its disruption.

%\subsection{Face Recognition Workflow}
\vspace{-0.2cm}
\subsection{\bf Breaking FR into Operational Stages}
\label{sec:dataflow}
\vspace{-0.1cm}

We now examine the FR operational pipeline and divide it into a set of {\em operational stages} to help frame our discussion of AFR tools. These operational stages correspond to specific subtasks in FR, which encompass {\em the five critical points of direct interaction between users and FR systems. }
Figure~\ref{fig:dataflow} depicts the five operational stages of a FR
system \adv~=
$\{\mathcal{G}, \mathcal{F}, \mathcal{C}, \mathbf{D}\}$. 
We discuss each stage below and revisit them as a framework to
analyze AFR tools in \S\ref{sec:stages}.

The overall operation of a FR system includes two phases: a 
{\em training \& enrollment} phase where the system builds (or acquires) a face feature
extractor and  creates a reference database
of known identities, and a  {\em run-time recognition}
phase where the FR system
identifies an unknown face.  As we show below,  the training \& enrollment
phase employs stage $\crc{1}$--$\crc{4}$, while the 
recognition phase employs stage $\crc{1}$, $\crc{2}$ and $\crc{5}$. 

\looseness=-1 \para{Stage $\crc{1}$: collecting face images.}  Face images primarily come from two
sources: online image scraping~\cite{clearview_explained} or
physically taking a photo of a person~\cite{perpetual,nist_fvr_test}.
We discuss sources of face images for FR systems in further detail in \S\ref{subsec:current}.

\looseness=-1 \para{Stage $\crc{2}$: preprocessing raw face images via
    $\mathcal{G}$.} Raw images obtained from stage $\crc{1}$ are
often poorly structured (\eg  varying face sizes, bystanders in
background). To make downstream tasks easier, \adv~ employs an image preprocessing engine $\mathcal{G}$ that uses face
detection (\eg automated face cropper~\cite{zhang2016joint}) to
remove background and extract each individual face, followed by a data normalization
process~\cite{arcface, cosface, facenet}.

\para{Stage $\crc{3}$:  training a feature extractor $\mathcal{F}$}.  The crucial element of DNN-based FR 
systems is the feature extractor $\mathcal{F}$ used to compute facial features from an image.  To
achieve accurate recognition, the computed feature vectors must be highly
similar for photos of the same person, but sufficiently
dissimilar across photos of different people. To enable this behavior, most existing
FR systems adopt the training methodology proposed by~\cite{facenet}
in 2015:  adding an {\em extra} loss function during $\mathcal{F}$ training to directly 
optimize for large separations between different faces in the feature
space. Followup works explore alternative loss 
functions and architectures to further improve the accuracy of
FR systems ({e.g.},~\cite{arcface,cosface,magface}). 

To maximize efficacy, $\mathcal{F}$ is generally trained on millions
of labeled face images. Extensive resources are required to
both collect and label a large face dataset and to actually train the
model. As a result, many FR practitioners, including
large companies~\cite{amazon_cus} and government
agencies~\cite{gao_2020,gao_2021}, opt to purchase or license a well-trained
feature extractor ({e.g.\/}~\cite{idemia, clearview, azure,
  amazon_rek, megvii, sensetime, yitu, cloudwalk}). We refer to images used in stage $\crc{3}$ as {\em training images}.

\looseness=-1\para{Stage $\crc{4}$: creating a reference database $\mathbf{D}$.}  FR systems need a large database
of known (labeled) faces in order to identify unknown (unlabeled) faces. As a result, FR operators
must build a reference database $\mathbf{D}$ of people they want to recognize, by first
collecting and preprocessing labeled face images of these individuals, and then 
passing them to $\mathcal{F}$ to obtain feature vectors. The
reference database holds the (feature vector, identity) pairs~\cite{
aws_rekognition, clearview_explained, megvii_explained}.  We refer to images used in stage $\crc{4}$ as {\em reference images}.

\para{Stage $\crc{5}$: recognize the face in a query image via $\mathcal{C}$.} At run-time, the FR system takes in and preprocesses (via $\mathcal{G}$)  a query image (i.e., an
unidentified face image),  extracts its
feature vector via $\mathcal{F}$, and queries the reference database $\mathbf{D}$ to locate a match (if any). If the feature
space distance ({{e.g.\/}, $L_2$ or cosine) of the query image is
  sufficiently close to a stored entry in $\mathbf{D}$, the system outputs a
  match. In this paper, we represent this process by the classifier $\mathcal{C}$.

\vspace{-0.1cm}
\subsection{FR Deployment and Data collection}
\vspace{-0.05cm}
\label{subsec:current}

In recent years, entities across the globe
have adopted and deployed FR systems for various applications. This wide adoption was triggered by
significant accuracy improvements of FR systems, largely due to new training
methods~\cite{facenet} and more powerful neural network
architectures~\cite{szegedy2016inception}. Deployment scenarios
  of these FR systems, along their data sources, have informed AFR
  tool development. Thus, to contextualize AFR proposals, we briefly
  examine how FR is used in the real world and from where its images
  (i.e., training/reference/query images) are drawn.

\para{Deployment scenarios.} Both public and private entities use FR
for a variety of purposes. We list some examples in
Table~\ref{tab:government_use}. Public (e.g.,  government-based) FR use cases range from criminal identification\footnote{Recently, police departments around the US have drawn fire
for their use of highly unregulated FR software like
Clearview.ai~\cite{clearview}.}\cite{garbagein}, civilian surveillance~\cite{uighur_surveill, megvii_uighur} and border control~\cite{cbp_facialrec}, to video game
use tracking~\cite{tencent_videogames} and COVID
lockdown enforcement~\cite{china_covid}. For a broader exploration of government uses of facial recognition, we refer the reader to~\cite{visual_capitalist}. Private entities have also integrated FR into their security and commerce pipelines. The most common private FR use cases are enhancing store or office security, but other examples abound (see Table~\ref{tab:government_use}).

  \para{Sources of face images.} The definitive source of images for deployed
  FR models is often unknown. Based on government reports and media articles,
  we outline some known sources of training, reference, and query images used
  today.

  {\em Training images} (used to train the feature extractor $\mathcal{F}$) often come from a mix of
  academic training datasets (\eg~\cite{msceleb,vggface2,facescrub,webface}), proprietary
  data, and public data scraped from social media accounts, according
  to a report of the US Government Accountability
  Office~\cite{gao_2020}.

  {\em Reference images} (used to create
  the reference database) generally come from the Internet (\eg social media), or government databases (\eg passport and driver license
  photos). Table~\ref{tab:labeled_data} shows a list of known
  reference image sources for some well-known FR operators.

  {\em Query images}, or faces to be identified by the FR system at
  run-time,  can come from both
  online and physical sources.  Some known sources include social media, police body cams,
  mug shots, corporate surveillance systems, state ID images, and
  passport photos~\cite{gao_2021}. After identification, query
  images are sometimes fed back into the reference database, either to enhance existing feature vectors or to create new ones. For example, US Customs and
  Border Patrol states that images of non-US travelers collected
  at US entry points are fed back into a larger DHS database as reference
  images~\cite{gao_tsa}. Similar techniques are used by several Chinese
  companies~\cite{china_commercial_fr, china_realestate_fr}.

\section{{\bf Anti-Facial Recognition: Motivation and Threat
    Model}}

\label{sec:threat}

%\subsection{Anti-FR Sentiment and

In this section, we discuss factors driving the development of
anti-facial recognition (AFR) tools, the threat model of those AFR
tools, and its practical implications. 

\begin{figure*}[t]
  \centering
  \includegraphics[width=0.95\textwidth]{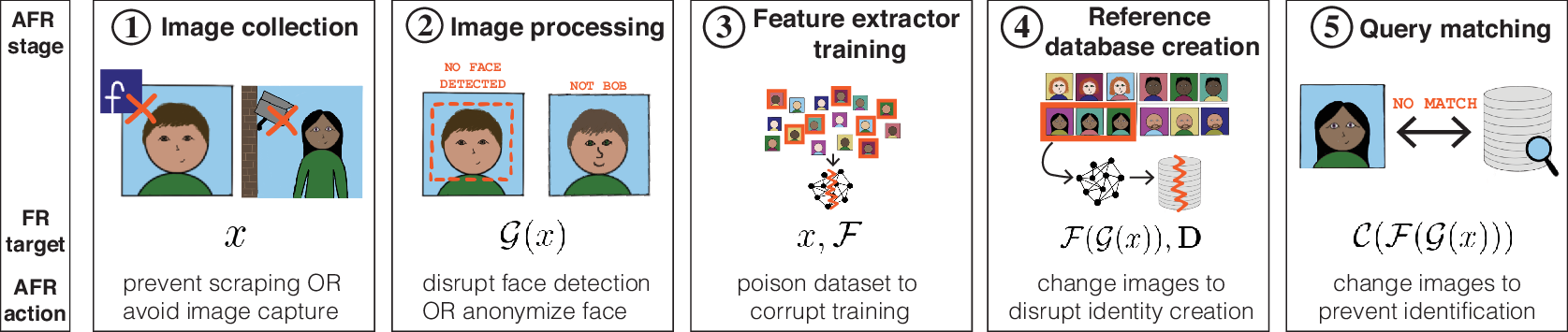}
  \vspace{-0.1in}
  \caption{Overview of our proposed stage-based framework for analyzing
    existing AFR proposals.  We list the five critical stages of
    facial recognition as discussed in \S\ref{sec:dataflow} and present AFR strategies per stage by the
    attack target, action, and desired effect.}
  \label{fig:threat_points}
\end{figure*}

\vspace{-0.2cm}
\subsection{\bf The Rise of AFR Tools}
\vspace{-0.2cm}

Numerous forces have coalesced to drive the recent trend in AFR tool development.
{\em First}, numerous reports about the provenance of images used in
commercial FR systems have raised significant privacy concerns.
The most infamous examples are Clearview.ai and PimEyes -- both
companies have scraped over \textit{3 billion images}
from social media sites to use in their FR
systems~\cite{hill_clearview} without user knowledge or consent.  {\em Second}, increased government use of
FR systems has caught the attention of citizens who have raised significant
concerns about the long-term effects of FR on privacy and freedom of expression~\cite{bigbrother, hongkong}. 
{\em Third}, multiple editorials have highlighted and discussed the demographic bias of existing FR systems, calling for a moratorium on
(or at least regulation of) the FR technology~\cite{ibm_ceo_letter, fr_nyt_opinion, atlantic_opinion}.

\looseness=-1 Consequently, public sentiment about FR is mixed and,
especially in western countries, trending negative~\cite{pewsurvey,
  ada2019beyond, steinacker2020facial, lai2021has}. This shift in
public opinion, combined with the forces noted above, has motivated researchers to create various AFR tools to counteract
unwanted FR systems.

\vspace{-0.2cm}
\subsection{\bf Threat Model of AFR}
\label{subsec:threat}
\vspace{-0.2cm}

AFR tools are used by a person \pr~to combat a FR system
\adv~\erev{= $\{\mathcal{G}, \mathcal{F}, \mathcal{C}, \mathbf{D}\}$}. In this context, \pr~takes the role of an
attacker and acts against \adv. \erev{\pr's goal is to prevent
  recognition by \adv, i.e., given an image $x_P$ of $P$, a successful AFR tool should cause \adv~to produce $C(F(G(x_P))) \ne P$.}

Proposed AFR tools generally make the following assumptions about each party:
%\noindent\rule{\linewidth}{0.4pt}
\begin{packed_itemize}
  \item \pr~has no special access to or authority
    over \adv, but wishes to evade
    unwanted identification by modifying or otherwise controlling their own face images. 
  \rev{
  \item \pr~wishes to avoid facial recognition, but also may wish for their images to remain useful for other purposes. For example, if \pr~posts a headshot on a personal website, they would like to ensure that the headshot not be scraped and used in a FR engine but also that their face remains recognizable to website visitors. Thus, \pr~prefers AFR tools which maximize AFR protection while minimizing image disruption. % Consequently, \pr~requires that AFR tools introduce only minimal distortion to images.
  }
  \item \adv's goal is to either create or
    maintain an accurate facial recognition operation. Furthermore,
    \adv~operates {\em at scale} and does not specifically target \pr~for identification.
  \end{packed_itemize}

  %\noindent\rule{\linewidth}{0.4pt}

  \para{Implications.} \erev{We also explore the real-world
    implications of the above threat model.} 
%  The threat model relies on assumptions, whose implications are explored below.

\vspace{3pt} \noindent (1) {\em Assuming AFR tools operate on images} --   Our 
study focuses exclusively on image-based AFR tools that a user
\pr~can deploy on their own. These image-based designs, \erev{which operate either directly on images or on systems that collect/process images}, dominate the current
set of AFR proposals.   On the other hand,  a user \pr~may, depending on the context,
be able to use other means (\eg legal action) to fight
unwanted facial recognition. 
\vspace{3pt}\noindent (2) {\em Assuming \adv~does not specifically target
  \pr~for recognition} --  Existing AFR tools are
designed to fight large-scale FR systems.  This is because,
from a practical standpoint, if system \adv~wishes to specifically
recognize a user \pr, there are much more efficient options than using
a general, large-scale FR system. Therefore, most current AFR tools are not
designed to withstand this level of scrutiny.  \erev{If \adv~makes a more targeted effort to identify \pr, such as hiring a private investigator, current AFR tools will likely fail.}

\rev{
\looseness=-1 \vspace{3pt}\noindent (3) {\em Assuming AFR tools minimize perturbations.} This study focuses on AFR tools which introduce {\em minimal perturbations} to images (as measured by $L_P$ norms). This decision is grounded in prior work showing that users are more likely to use privacy-preserving tools with minimal overhead~\cite{faceoff,coopamootoo2020usage,wright2002privacy}. AFR tools which do not seek to minimize perturbations are not addressed in this work. This is a limitation of our work, and future work should consider AFR tools which use metrics beyond $L_P$ norms to measure image distortion.
}

%\section{{\bf A Stage-based Framework for Systematizing AFR Tools}}
\vspace{-0.4cm}
\section{A Stage-Based AFR Framework}
%Proposals}}
\vspace{-0.2cm}
\label{sec:stages}

We now discuss and analyze existing AFR proposals.  To do so, we propose and use a {\em stage-based} framework to categorize AFR
strategies. As discussed in \S\ref{sec:back}, a FR system \adv~=
  $\{\mathcal{G}, \mathcal{F}, \mathcal{C}, \mathbf{D}\}$ operates in
  $5$ distinct stages that correspond real-world actions (e.g. image
  capture, pre-processing, feature extraction, etc.). In each stage,
  \adv~interfaces with $P$ and the broader world as it collects,
  processes, and uses image data. Such interfaces represent possible
  points at which $P$ can act against \adv.
  Specifically, an AFR tool can attack the component(s) of
  \adv~relevant to any of the $5$ stages, including images $x$,
  preprocessor $\mathcal{G}$, feature extractor $\mathcal{F}$,
  reference database $\mathbf{D}$, and classifier $\mathcal{C}$.  With
  this in mind, Figure~\ref{fig:threat_points} demonstrates the attack
  actions and goals when AFR tools target each FR stage.

\begin{table*}[t]
\centering
\resizebox{0.99\textwidth}{!}{

\begin{tabular}{|c|c|c|c|c|c|c|c|} 
 \hline

  \multicolumn{1}{|c|}{\multirow{3}{*}{\begin{tabular}[c]{@{}c@{}}\textbf{AFR}\\\textbf{system}\end{tabular}}} & \multicolumn{1}{c|}{\multirow{3}{*}{\begin{tabular}[c]{@{}c@{}}\textbf{Year}\\\textbf{released}\end{tabular}}} & \multicolumn{1}{c|}{\multirow{3}{*}{\begin{tabular}[c]{@{}c@{}}\textbf{Stage}\\\textbf{targeted}\end{tabular}}} & \multicolumn{5}{c|}{\textbf{Attack scenario}} \\ 
\cline{4-8} \multicolumn{1}{|c|}{} & \multicolumn{1}{c|}{} & \multicolumn{1}{c|}{} & \multicolumn{1}{c|}{\begin{tabular}[c]{@{}c@{}}\textit{\pr's knowledge}\\\textit{of \adv}\end{tabular}} & \multicolumn{1}{c|}{\begin{tabular}[c]{@{}c@{}}\textit{\pr's operating}\\\textit{context}\end{tabular}} & \multicolumn{1}{c|}{\begin{tabular}[c]{@{}c@{}}\textit{Targeted/}\\\textit{Untargeted}\end{tabular}} & \multicolumn{1}{c|}{\begin{tabular}[c]{@{}c@{}}\textit{Tested on}\\\textit{real-world FR}\end{tabular}} & Unique Property \\ 
  \hline
%[98-102]
Anti-scraping [98-102] & 2021 & $\crc{1}$ &  - & Digital & UT & - &
Prevent large-scale image scraping
 \\  
Data Leverage~\cite{dataleverage} & 2021 & $\crc{1}$  &  -
& Digital & UT &- & Withholds data to prevent collection. \\\hline \hline

CVDazzle~\cite{cvdazzle} & 2010 & $\crcs{2a}$ &  WB & Physical & UT & - & \shawn{Make-up} \\ 
 Xu et al.~\cite{xu2020adversarial} & 2020 & $\crcs{2a}$ &  BB & Physical & UT & YOLOv2 & \shawn{Adversarial patch on T-shirts} \\
Wu et al.~\cite{wu2020making} & 2020 & $\crcs{2a}$ &  Both & Physical & UT & YOLOv2 & \shawn{Adversarial patch on T-shirts} \\ 
Zolfi et al.~\cite{zolfi2020translucent} & 2020 & $\crcs{2a}$ &  BB & Physical & UT & YOLOv5 & \shawn{Stickers on camera lens that blur vision} \\
SocialGuard~\cite{xue2020socialguard} & 2020 & $\crcs{2a}$ &  WB & Digital & UT & - & \shawn{Adversarial perturbation on face detectors} \\ 
Hu et al.~\cite{chen2021naturalistic} & 2021 & $\crcs{2a}$ &  WB & Digital & UT & - & Adversarial patch on object detectors \\ 
Treu et al.~\cite{treu2021fashion} & 2021 & $\crcs{2a}$ &  BB & Digital & UT & - & \shawn{Adversarial clothing on face detectors} \\
DeepPrivacy~\cite{deepprivacy}  & 2019 & $\crcs{2b}$ &  BB & Digital & UT & - & \shawn{GAN-based face blurring (perceptible)} \\
IdentityDP~\cite{identitydp}  & 2021 & $\crcs{2b}$ &   BB & Digital & UT & AZ & \shawn{GAN-based face blurring (perceptible)} \\ 
DeepBlur~\cite{deepblur} & 2021 & $\crcs{2b}$ &  BB & Digital & UT & AZ, F++ & \shawn{GAN-based face blurring (perceptible)}  \\
Yang et al~\cite{yang2021systematical} & 2021 & $\crcs{2b}$ & BB & Digital & UT & - & \shawn{GAN-based face blurring (imperceptible)} \\
  
\hline \hline

%WitchesBrew~\cite{witches} & 2020  & Training & Poisoning &  & Digital & &
%& \\
Evtimov et al.~\cite{evtimov2021disrupting} & 2021 & $\crc{3}$ &  BB & Digital &UT &  - & \shawn{Data poison by modifying entire dataset} \\ 
Huang et al.~\cite{unlearnable} & 2021 & $\crc{3}$ &  BB & Digital & UT & - & \shawn{Data poison by user coordination} \\ 
\erev{Fu et al.~\cite{fu2021robust}} & 2021 & $\crc{3}$ &  BB & Digital & UT & - & \shawn{Data poison by unlearnable data} \\ \hline \hline

%FishyFace~\cite{fishy}  & 2018 & $\crc{4}$ &  WB & Digital & T & - & Attacks SVM training. \\
Fawkes~\cite{fawkes}  & 2020  & $\crc{4}$  &  Both & Digital & UT & AR, AZ, F++ & Corrupts features of faces \\
FoggySight~\cite{foggy} & 2021 & $\crc{4}$ &   Both & Digital & UT & AZ & Collectively corrupts features of faces \\
LowKey~\cite{lowkey} & 2021 & $\crc{4}$ &  BB & Digital & UT & AR,
                                                                 AZ  &
Corrupts features of faces \\ \hline \hline
  
Feng et al.~\cite{feng2013facilitating} & 2013 & $\crc{5}$ &  BB & Physical & UT & - &  Make-up \\
Sharif et al.~\cite{accessorize} & 2016 & $\crc{5}$ &  Both & Both & Both & F++ & Adversarial patch on wearable accessories \\
Dabouei et al.~\cite{fastperturb} & 2018 &  $\crc{5}$ &  WB & Digital & UT & - & Adversarial attack distorts face landmarks. \\
Zhou et al.~\cite{invisiblelight} & 2018 & $\crc{5}$ &  WB & Physical & Both & - & Projected adversarial IR patterns \\
Dong et al.~\cite{efficientblackbox} & 2019 & $\crc{5}$ &  BB & Digital & T & TN & Black-box adversarial perturbation. \\ 
Zhu et al.~\cite{makeupevasion} & 2019 & $\crc{5}$ &  Both & Digital & Both & - & Adds eye makeup with GAN. \\
AdvHat~\cite{advhat} & 2019 & $\crc{5}$ &  WB & Physical & UT & - & Printed sticker on hat. \\  
AdvFaces~\cite{advfaces} & 2019 & $\crc{5}$ &  BB & Digital & Both & - & GAN-based adversarial attack. \\
VLA~\cite{vla} & 2019 & $\crc{5}$ &  BB & Physical & Both & - & Projected light patterns \\
Nguyen et al.~\cite{adversariallight} & 2020 & $\crc{5}$ &  Both & Physical & Both & ? & Projected light patterns \\
Browne et al.~\cite{cameraadversaria} & 2020 & $\crc{5}$ &  BB & Digital & UT & - & Universal adversarial perturbation \\ 
Cilloni et al.~\cite{preventing} & 2020 & $\crc{5}$ &  WB & Digital & UT & - & Corrupts features of faces \\ 

Face-Off~\cite{faceoff}& 2020 & $\crc{5}$ &  BB & Digital & Both & AR, AZ,
                                                             F++ &
Study on user perception on perturbation levels. \\
  Singh et al.~\cite{singh2021brightness} & 2021 & $\crc{5}$ & WB &  Digital
& UT & - & Brightness-agnostic adversarial perturbations \\

Yang et al~\cite{yang2021towards} & 2021 & $\crc{5}$ & BB & Digital & UT &TN & Corrupts features of faces\\
  
\hline
\end{tabular}}
\vspace{0.1cm}
\caption{Taxonomy of proposed AFR tools. ``BB/WB'' = Black Box, White
  Box.``UT, T'' = Untargeted, Targeted. ``AR, AZ, F++, TN'' = Amazon Rekognition, Microsoft Azure Face Recognition, Megvii's Face++, Tencent Face Recognition.}
\label{tab:taxonomy} \vspace{-0.1in}
\end{table*}

\vspace{-0.2cm}
\subsection{\bf AFR Strategies per Stage}
\vspace{-0.2cm}

Since the five FR stages {\bf $\crc{1}$}--{\bf $\crc{5}$} encompass the points of
direct interaction between \pr~and \adv, they naturally cover the  
points of attack employed by existing AFR proposals.  Next we briefly describe the general strategies used by AFR tools targeting each FR
stage. 

\para{Attacking $\crc{1}$.} 
In the image collection stage, labeled and/or unlabeled images \erev{$x$} are collected for
use by \adv, either by physically taking photos or scraping online
images. \rev{Labelled images can be used as training or reference images to build a FR system, while unlabelled images can be used as query images.} When targeting this stage, AFR tools focus on disrupting the data collection process
to prevent \adv~from acquiring usable face images $x_P$ of \pr. 

\para{Attacking $\crc{2}$.}
In the second stage, \adv~uses $\mathcal{G}$ to pre-process collected face images using a series of digital transformations, {e.g.}, face detection, background cropping, and normalization. AFR tools deployed at this stage target $\mathcal{G}$ to render the processed images unusable, either by
breaking the preprocessing functions ({e.g.}, preventing faces from
being detected) by injecting noise and artifacts onto the
images or removing \pr's identity information from the images. \rev{We denote these different actions as $\crcs{2a}$ and $\crcs{2b}$, respectively.}

\looseness=-1 \para{Attacking $\crc{3}$.} Since stage $\crc{3}$ is dedicated to training \adv's feature
extractor $\mathcal{F}$, AFR tools targeting this stage seek to degrade the accuracy of $\mathcal{F}$ by poisoning its training images.

\para{Attacking $\crc{4}$.}
To create the reference database \erev{used by classifier $\mathcal{C}$}, labeled reference images are passed
through \erev{$\mathcal{F}$} to create their feature vectors.  AFR
tools targeting this stage attempt to corrupt the feature vectors created for
\pr's reference images so that the database holds a ``wrong'' feature vector
of \pr, \erev{and $\mathcal{C}$ fails}.

\para{Attacking $\crc{5}$.}
% (Query Matching).} 
In the query matching stage,  AFR tools seek to prevent
\erev{classifier $\mathcal{C}$} from accurately matching query
image \erev{$x_P$'s} feature vector and \pr's feature vectors stored in
\adv's reference database. This is generally achieved by perturbing
(or modifying) the query image \erev{to change its feature vector and thwart $\mathcal{C}$}. 

\vspace{-0.2cm}
\subsection{\bf Taxonomy of Existing AFR Proposals} 
\label{subsec:litreview}
\vspace{-0.2cm}

Using our stage-based analysis framework, we now present a comprehensive taxonomy of existing AFR
proposals in Table~\ref{tab:taxonomy}.  In this list, we categorize existing AFR proposals by  the
year of release, the individual FR stage they target, and the attack
scenario.  We further break down the attack scenario by \pr's knowledge of \adv~(white box or
black box\footnote{{\em White box} means \pr~has full access to \adv's FR system (including feature extractor parameters) and uses this knowledge to guide their
AFR protection. {\em Black box} means \pr~lacks such access and
knowledge.}), the AFR deployment context (physical or digital),  whether the attack is targeted or untargeted\footnote{A {\em targeted} attack causes the FR system to identify \pr~as a specific, incorrect person (e.g. a famous politician). An {\em
untargeted} attack means that \pr~is misclassified, but not as a specific person.}, whether the AFR tool has been tested against real-world FR systems, and any unique or notable features of the
AFR tool. \rev{We note that the majority of AFR tool users may not care for or need a {\em targeted} AFR misclassification result, but we include targeted attacks for completeness, as they represent the most user-controlled version of an AFR tool.}

There is a significant imbalance of AFR tools targeting different stages. Stage $\crc{2}$ and $\crc{5}$ have \htedit{attracted} the most number of AFR proposals, likely due to 
the popularity of adversarial perturbation research. We also notice that 7 out of 30
proposals assume a ``white-box'' access to \adv's FR pipeline, which is often unrealistic in
practice. Finally, only 12 out of the 30 proposals have tested the AFR effectiveness against
at least one real-world FR system. Overall, Table~\ref{tab:taxonomy} serves as a comprehensive summary
of current AFR proposals, which we will refer to throughout the paper. 

\erev{
\para{Adversarial ML and AFR.} A significant portion of AFR tools
listed in Table~\ref{tab:taxonomy}, e.g., those targeting $\crcs{2a}$
and $\crc{3}-\crc{5}$,  apply adversarial machine learning (AML)
techniques like poisoning or evasion attacks to thwart
\adv. Consequently, a significant portion of this paper is devoted to
discussing the pros and cons of AML-based approaches to AFR.  On the
other hand,  since AFR tools targeting stage $\crc{1}$ and $\crc{2b}$ are
inherently non-AML based,  our analysis is not limited to only
AML-enabled AFR tools.  Since each of the five FR stages represents a 
viable attack vector for AFR tools,  our analysis covers AFR tools
targeting any stage.

\vspace{-0.1cm}
\subsection{{\bf Roadmap of Our Analysis}}
\vspace{-0.1cm}

Using the stage-based framework, we conduct a detailed analysis of
existing AFR proposals.  First,  we discuss in greater detail how
existing AFR proposals attack each of the five stages
(\S\ref{sec:collection} -- \S\ref{sec:evade}).  In each section, we
describe the goals of \adv~and \pr,  the challenges of targeting
this particular stage, the existing proposals, and a summary of the
progress made in this direction. Next, we conduct a meta-analysis of 
AFR strategies across the five stages (\S\ref{sec:tradeoffs}), and
discuss what we see as the major technical and broader social/ethical
challenges facing future AFR development  (\S\ref{sec:future}). }

\section{{\bf Attacking $\crc{1}$ to Disrupt Data Collection}}
\label{sec:collection}
We start by examining AFR methods that allow \pr~to attack
\adv~by disrupting the process of face data collection.

\looseness=-1 \para{ Goals and Challenges.} In this data collection stage, \adv's goal is to obtain usable face images \rev{$x$} from
online or physical sources.  Often, \adv~aims to collect high quality images of millions or billions of people (\eg
Clearview.ai~\cite{hill_clearview}).   \adv~uses labeled images  to build the
reference database and/or train the feature extractor. 
By using AFR tools, \pr's goal is to prevent their face images \rev{$x_P$} from being collected for use by \adv, either
as (labeled) reference images or (unlabeled) query images. They can apply online or physical evasion/disruption techniques to do so.
%\end{packed_itemize}

The key challenges facing AFR proposals targeting this stage are that (1)
they need to be aware of and adapt to \adv~who continues to 
innovate techniques for data collection; and (2) cameras are widely deployed in the
real-world, making it challenging to avoid image capture by
cameras. 

\vspace{-0.1cm}
\erev{
\subsection{\bf Current Solutions}
\label{sec:download}
}
\vspace{-0.1cm}
Face images can come from two sources: scraping online images or
physically capturing faces using cameras. \erev{ Image scraping refers to the collection of images posted online that were captured by someone who is not the data collector (e.g. camera is operated by the subject, subject’s friend, etc). Picture-taking refers to images taken directly by the data collector.} Thus we divide AFR tools acting at this stage into two categories: preventing scraping and preventing capture. 
 
\para{Preventing Online Image Scraping.} A large portion of face images used in today's FR systems are scraped from online social media platforms. Thus, an effective way to stop \adv~is to prevent web scraping.  While each single user can try their best to limit their online footprint, most AFR methods require an online platform (\eg Flickr) or outside help.   

\begin{packed_itemize}
  \item {\em Anti-scraping by online platforms.}  Anti-scraping techniques have been widely studied in
the security
community~\cite{wang2013you,gold2017robots,parikh2018detection,jawad2017detection,haque2015anti}. Techniques
such as rate limits, data limits, ML-based scraping detection are already used by online
platforms~\cite{fbreport}. However, a significant portion of
scraping still goes undetected as scrapers develop more sophisticated tools to
bypass detection~\cite{fbreport}.  
   \item {\em Data leverage by users. } \pr~could try to prevent \adv~from
   collecting their online images by withholding them. Recent works
   propose the concept of ``data leverage'' where users of online platforms work collectively to withhold data or control how their data is used by tech
   companies~\cite{dataleverage, vincent2021can, vincent2019data}. While
   not specifically aimed at facial recognition, these proposals offer
   alternative models for online engagement while protecting user
   data. 
\end{packed_itemize}

\para{Avoiding Image Capture.} Ordinary civilians can already use smartphones to take high-quality photos of anyone at any moment.  These photos could be collected and used by facial recognition systems like PimEyes~\cite{pimeyes}.  Furthermore, face
photos taken by on-street surveillance cameras are increasingly used by commercial or government
facial recognition systems~\cite{intel, apple, walmart, perpetual,
  facesurveillance}, especially in major metropolitan areas and inside stores. Today's proposals for avoiding image capture come from both research
community and activists (e.g. protesters and artists)
concerned about surveillance.  They fall into two categories:
{\em hiding faces from cameras} and {\em disrupting camera operation}.

\begin{packed_itemize}
  \item {\em Face hiding.} People can wear clothes, hats, masks, or move
  their head to prevent (usable) facial image being captured by
  cameras. Notably, during the June 2020 wave of protests in the US,
  nonprofit organizations compiled a ``tech toolkit'' to help privacy-conscious protesters obfuscate their faces
  from cameras and avoid identification~\cite{amnesty}; in 
  late 2020, a Chinese artist used a map of on-street surveillance
  cameras to successfully guide others to evade identification by positioning their
  head/body ``away'' from those cameras~\cite{china_artist}. 
\item {\em Camera disruption.} Without physically breaking cameras, human
users can prevent cameras from capturing (usable) images by simply
shining laser lights at them~\cite{hongkong}. Other methods include covering cameras with fabric or stickers. 
\end{packed_itemize}

\vspace{-0.2cm}
\subsection{{\bf Discussion of Stage $\crc{1}$ Solutions }}
\vspace{-0.2cm}

\rev{

\looseness=-1 \para{Privacy/Utility Trade-offs. } Evading data collection requires both fine-grained
control over one's online identity and awareness of when/how
pictures are being taken, making it difficult for users to deploy these tools without significantly limiting either their online or physical activities. Furthermore, there are cases where evading data collection is simply impossible, \ie mandatory pictures posted on an employer's website. Anti-scraping tools can also decrease the utility of the service provider, as such tools can have false positives and a high deployment cost. 
}

\looseness=-1 \para{Summary of Progress. } Existing AFR proposals against stage $\crc{1}$ make headway in
addressing key challenges. Adaptive anti-scraping techniques
like~\cite{fbreport} definitely raise the bar for
attackers. Furthermore, anti-data collection methods
like~\cite{china_artist} have shown that it is possible, with careful
action, to evade image capture even in robust surveillance
systems. Future AFR development against stage
$\crc{1}$ can seek to improve both data controls and camera awareness
by individual users. 
\vspace{-0.1cm}
\section{\bf Attacking $\crc{2}$ to Disrupt Face Pre-processing }
\label{sec:preprocess}
\vspace{-0.5cm}

\looseness=-1 In stage $\crc{2}$, \adv~processes raw face images \erev{with $\mathcal{G}$} to facilitate further operations in stages $\crc{3}$, $\crc{4}$, and $\crc{5}$. AFR proposals targeting this stage seek to disrupt $\mathcal{G}$ \rev{by transforming an image $x$ into $x'$} such that processed face images \rev{$\mathcal{G}(x')$} are ``unusable'' by subsequent FR stages.

\para{\bf Goals and Challenges.} \adv's goal is to \erev{use $\mathcal{G}$} to obtain well-structured face images from many raw images. \pr's goal is to either prevent their face being detected/extracted from raw images \erev{by $\mathcal{G}$} or to anonymize their face in these images. 

The main challenge for AFR proposals targeting stage $\crc{2}$ is how to achieve {\em anonymization without distortion}.  That is,  when modifying P's images to either evade $\mathcal{G}$'s face detection or to remove identity information, the modified images should still resemble $P$ to remain {\em useful} to $P$. An additional challenge is overcoming (adaptive) defenses deployed by \adv~to protect $\mathcal{G}$.

\vspace{-0.1cm}
\subsection{\bf Current Solutions}
\vspace{-0.1cm}

\para{Preventing Face Detection.} Face detection extracts well-centered headshots from raw images. The commonly used face detection  systems~\cite{zhang2016joint} rely on DNNs to accurately infer the location of faces in an image.  \rev{To prevent effective face detection/extraction, the AFR goal is to produce an adversarial $x'$ such that $\mathcal{G}(x') = z$, where $z$ is a useless result that cannot be passed on to $\mathcal{F}$.}
To create \rev{$x'$},  existing AFR tools leverage ``adversarial perturbations'' against DNN models.  Adversarial perturbations are a well-studied phenomenon in the field of adversarial machine learning. These carefully crafted, pixel-based perturbations, when added to an image, can cause DNNs to produce wrong classification results ({e.g.,\/}~\cite{madry2017towards,cwattack,chen2018ead,bao2020sparse}). Perturbations are generated using an iterative optimization procedure that maximizes the likelihood of model misbehavior while minimizing perturbation visibility. The generation procedure varies depending on \pr's knowledge of \adv~(e.g. white vs. black box, see Table~\ref{tab:taxonomy}).

AFR tools using adversarial perturbations can be subdivided based on how the perturbation is added to images. They
can be directly added to digital images if \pr~has direct
access to these images or fabricated as physical objects that
\pr~can wear ({e.g.,\/} an adversarial T-shirt) or place on
cameras.

\begin{packed_itemize}
  \item {\em Directly modifying digital images.} Using AFR tools, users
  who post images online can directly add adversarial perturbations to these images
  before posting them (\eg~\cite{xue2020socialguard,
    treu2021fashion,chen2021naturalistic}). Properly
  perturbed images cannot be used by FR systems to extract any face
  information. 
  \item {\em Wearing custom designed physical objects.}  Often
users do not have access to face images to modify them. An 
alternative way to ``inject'' adversarial perturbations into images is
to carry or wear a physical object so that any camera taking a photo of the user will
also capture a version of the adversarial perturbation.  Along these lines, prior works have successfully translated face-detection-evading adversarial
perturbations into makeup~\cite{cvdazzle, amnesty},
t-shirts~\cite{xu2020adversarial, wu2020making}, or stickers. 
  \item {\em Placing a sticker on cameras.} An orthogonal approach involves transforming the adversarial
perturbation into a translucent sticker that can be placed over a
camera lens. This sticker imperceptibly modifies images taken by the camera
to prevent people and faces from being detected in those images~\cite{zolfi2020translucent}. 
\end{packed_itemize}

\looseness=-1 \para{Anonymizing Faces $\crcs{2b}$.} \pr~can also {\em anonymize} their face images to remove identity
information. \rev{In this setting, \pr~creates an $x'$ such $\mathcal{G}(x') \ne \mathcal{G}(x)$, i.e. the result after processing is still usable but represents a fake identity.} {\em Physical anonymization} can be easily achieved by wearing masks, hats, makeup, etc,  which overlaps with ``avoiding image capture'' in $\crcs{1}$ discussed in \S\ref{sec:collection}. Leaving proposals for {\em digital anonymization} use generative adversarial networks (GANs)~\cite{gans} and differential
privacy~\cite{dwork2014algorithmic}. Several such proposals use GANs to transform face images into latent space vectors, modify those vectors to remove identity information, and reconstruct the images from the modified vectors~\cite{deepprivacy, deepblur, yang2021systematical}. The modified faces still look human but are anonymized to prevent accurate identification.  Another proposal, IdentityDP~\cite{identitydp}, uses similar techniques but also claim to provide differentially private identity protection.

\vspace{-0.2cm}
\subsection{{\bf Discussion of Stage $\crc{2}$ Solutions }}
\vspace{-0.2cm}

\rev{
\para{Privacy/Utility Trade-offs. } Many stage $\crcs{2a}$ proposals address the ``usability'' challenge by formatting adversarial patches against $\mathcal{G}$ as wearable clothing/objects. However, wearing this special clothing, 
which can appear bizarre, may not be desirable for the 
average person. Current proposals against $\crcs{2b}$ provide anonymity but tend to produce anonymized faces that do not resemble the original face, with significantly altered shape, skin tone, hair color, etc. These images lack many functionalities of traditional images, e.g. image sharing, preserving memories, etc. 
}

\para{Summary of Progress. } Many AFR proposals targeting $\crcs{2a}$ have been tested against real-world object detectors like YOLOv2, demonstrating their real-world efficacy. However, several defenses against these patches have emerged recently (e.g.,~\cite{liang2021we, liu2021segment, xiang2021detectorguard}),  although only one~\cite{liang2021we} has been tested against physical adversarial patches like the ones used by AFR tools~\cite{xu2020adversarial, wu2020making}. Further work is needed to determine if AFR proposals against $\crcs{2a}$ can resist these defenses that \adv~can use to protect $\mathcal{G}$. 

\vspace{-0.1cm}
\section{\bf Attacking $\crc{3}$ to Corrupt Feature Extractor}
\label{sec:poisongal}
\vspace{-0.1cm}

All FR systems require an effective feature extractor \erev{$\mathcal{F}$} to distinguish between faces. AFR proposals attacking stage $\crc{3}$ focus on corrupting the training of \erev{$\mathcal{F}$} to produce an unusable extractor \rev{$\mathcal{F}'$}.

\para{\bf Goals and Challenges.} Here, \adv's goal is to train a
high-quality feature extractor $\mathcal{F}$ using available training
data, so that faces can be accurately identified by their feature
vectors extracted by $\mathcal{F}$.  Thus \pr's goal is to prevent \adv~from training an effective
\erev{$\mathcal{F}$ by corrupting the training data}.

There are two key challenges facing AFR proposal targeting stage
$\crc{3}$. The first is minimizing the distortion to training face images introduced
by the corruption process while maintaining the corruption
efficacy. The second is corrupting $\mathcal{F}$'s training without requiring full 
dataset control. 

\vspace{-0.2cm}
\subsection{\bf Current Solutions}
\vspace{-0.2cm}

Data poisoning is a well-studied technique in the field of adversarial machine
learning. By manipulating the training data of a DNN model, an
external party can negatively impact the model's training~\cite{gu2017badnets,chen2017targeted,liu2018trojaning,zhu2019transferable,shafahi2018poison}. Poisoned
models can exhibit a variety of (mis)behaviors, from incorrect
classification of specific inputs to complete model failure.  Existing
AFR proposals focus on the latter. 

% In the
% stage $\crc{3}$ AFR context, poisoning proposals focus on the latter
% scenario. 

\para{Making training data unlearnable.} By injecting specially
crafted noise on training data, recent works~\cite{unlearnable,fu2021robust} render data ``unlearnable'' by a DNN model. This noise
misleads the model into thinking that data have already been learned, thwarting necessary parameter updates. When a user submits their ``unlearnable'' face images as a training image for the \erev{$\mathcal{F}$},  the extractor will not learn anything to improve its performance. Training an effective \erev{$\mathcal{F}$} requires millions or even billions of face images~\cite{arcface, cosface, facenet}, and with a sufficient number of unlearning training examples, \erev{$\mathcal{F}$} will not meet the accuracy level required for practical deployment. 

\para{Adding adversarial shortcuts.} A related proposal from Evtimov \etal~\cite{evtimov2021disrupting} injects \emph{adversarial shortcuts} into the dataset. Models trained on this data overfit to the shortcut and fail to learn the meaningful semantic features of the data. Now the trained extractor model has a distorted understanding of the feature space, \htedit{it cannot produce high quality feature vectors required for accurate face recognition.}

\vspace{-0.2cm}
\subsection{{\bf Discussion of Stage $\crc{3}$ Solutions }}

\vspace{-0.2cm}

\rev{
\para{Privacy/Utility Trade-offs. } The biggest utility drawback of stage $\crc{3}$ proposals is that they require
significant effort to corrupt the training dataset. Most proposals require that $P$ control much of 
the training data to render $\mathcal{F}$ unusable. Such high levels of control would prohibit 
individual users from using these AFR tools. In addition, \adv~ can discard a corrupted dataset and use other data sources to train their model, once they discover the presence of corruption. 

}

\looseness=-1\para{Summary of Progress. } Despite utility challenges, existing proposals have shown that, with a sufficient level of dataset
control, it is possible to render $\mathcal{F}$
unusable by adding minimally visible perturbations onto training face images. For example, AFR tools based on adversarial shortcuts~\cite{evtimov2021disrupting} are effective  when they can corrupt the entire training dataset. Others~\cite{unlearnable} can reduce $\mathcal{F}$'s accuracy on specific classes in a FR model by $16\%$. 

\section{\bf Attacking $\crc{4}$ to Corrupt Database}
\label{sec:poison}
In stage $\crc{4}$, \adv~uses \erev{$\mathcal{F}$} to create a reference
database $\mathbf{D}$ of labeled face feature vectors that will facilitate identification of unidentified faces. AFR tools targeting this stage seek to fill $\mathbf{D}$ with incorrect face/label mappings, so that \adv's classifier $\mathcal{C}$ cannot identify \pr's query images as \pr. \rev{Thus, when a true image $x_P$ is presented to \adv's system for identification, the corrupted database $\mathbf{D}'$ produces $\mathcal{C}(\mathcal{F}(\mathcal{G}(x_P)), \mathbf{D}') = I$}, where $I$ is an incorrect identity, $I \ne P$.

\para{\bf Goals and Challenges.} In this stage, \adv's goal is to create a reference database containing
accurate copies of feature vectors (\erev{produced by $\mathcal{F}$}) of people \adv~wishes to recognize. \pr's goal is to prevent \adv's feature extractor $\mathcal{F}$ from creating an accurate feature vector which \erev{$\mathcal{C}$ can match to} query images of \pr.  Note that this can also be achieved by corrupting the training data/process of $\mathcal{F}$ in stage $\crc{3}$, as discussed in \S\ref{sec:poisongal}.  Here, we differentiate from \S\ref{sec:poisongal} by assuming that $\mathcal{F}$ is a well-trained feature extractor. Thus, \pr~attacks \adv~by modifying/manipulating the reference images of \pr~that \adv~uses to create its reference database.

Stage $\crc{4}$ based AFR tools must first address the base case challenge of modifying \pr's reference images to produce incorrect feature vectors while minimizing the distortion of those images. They also face two advanced challenges. First, they must maintain high performance when \adv~has some original, unmodified face images of \pr~ already enrolled in its database $\mathbf{D}$. Second, protection must persist when \pr~makes incorrect assumptions about \adv's system, especially its extractor $\mathcal{F}$ and classifier $\mathcal{C}$, or when \adv~adapts.

\vspace{-0.1cm}
\subsection{\bf Current Solutions}
\vspace{-0.1cm}

Existing AFR proposals in this category focus on poisoning feature
vectors before they are stored in $\mathbf{D}$. The specific poisoning techniques depend on the underlying assumptions about \adv's classifier \erev{$\mathcal{C}$}.

\para{Assuming $\mathcal{C}$ uses classification-based matching.} A recent AFR proposal, Fawkes~\cite{fawkes},  assumes $\mathcal{C}$ is a shallow classification layer added to $\mathcal{F}$. Fawkes seeks to corrupt the final classification output by ``cloaking'' (or poisoning) reference images of \pr, \ie shifting their feature vectors away from the correct representation by adding imperceptible perturbations to \pr's reference images~\cite{fawkes}.  When $\mathcal{C}$ is trained on these shifted feature vectors, \adv~will learn to associate incorrect feature spaces with \pr's identity, producing wrong matches for \pr's (uncloaked) query images at run-time.

\looseness=-1 \para{Assuming \erev{$\mathcal{C}$} uses nearest neighbor-based matching.} Two other AFR proposals, LowKey~\cite{lowkey} and FoggySight~\cite{foggy}, assume \erev{$\mathcal{C}$} is a K-nearest neighbors algorithm.  LowKey~\cite{lowkey} adds digital adversarial perturbations to change the feature representation of \pr's reference images (similar to Fawkes). These perturbed images create a reference feature vector for \pr~that is different from those of \pr's run-time query images, thus \erev{thwarting $\mathcal{C}$}.  FoggySight~\cite{foggy} takes a community-driven approach, where users modify their
images to protect others. These collective modifications flood the top-K matching set for a specific user with incorrect feature vectors, drowning out the correct feature vector.

\rev{
\vspace{-0.2cm}
\subsection{\bf Discussion of Stage $\crc{4}$ Solutions}
\vspace{-0.2cm}

\para{Privacy/Utility Trade-offs.} Most Stage $\crc{4}$ proposals add perturbations directly to images. Several proposals discuss how stronger (more visible) perturbations yield stronger AFR protection (i.e.~\cite{fawkes, lowkey}). Visible perturbations may lower the utility of protected images, especially if they are meant to be posted on social media sites. More advanced optimization techniques may help reduce perturbation size at stronger protection levels, but this visibility/protection trade-off seems inevitable.

\para{Summary of Progress.} It is encouraging that all proposals listed in the above analysis have demonstrated success on the key task of corrupting feature vectors of \pr, i.e., the base case.  Overcoming the two advanced challenges discussed remains an area for future work. Some proposals provide limited protection when \adv~has obtained original, unmodified feature vectors of \pr~(e.g.~\cite{fawkes}), but not all proposals have considered this possibility. Second, all existing proposals assume knowledge of $\mathcal{C}$ and/or $\mathcal{F}$, necessitating further work to determine how/if incorrect assumptions of \adv~would affect AFR  performance. A final challenge is evaluating the long-term robustness of stage $\crc{4}$-based AFR mechanisms against an adaptive \adv. Recent work (discussed in \S\ref{sec:future}) suggests that a continuously adapting \adv~may always (or eventually) ``win'' against AFRs targeting $\crc{4}$. 

}

%\vspace{-0.2cm}
\section{\bf Attacking $\crc{5}$ to Evade Identification}
%\vspace{-0.2cm}
\label{sec:evade}
The final set of AFR tools aims to prevent run-time query image identification \erev{by \adv's classifier $\mathcal{C}$}, \rev{by producing distorted images $x_P'$ that mislead the final classifier outcome, e.g. $\mathcal{C}(\mathcal{F}(\mathcal{G}(x_P')), \mathbf{D}) = I \ne P$}. These methods can provide one-time protection for users who believe their images are already enrolled in $\mathbf{D}$. Furthermore, since labeled query images can also be added to the reference database, using these AFR tools at run-time can also help poison the reference feature vectors (see \S\ref{sec:poison}). However, current AFR proposals targeting this stage focus strictly on evasion and do not consider this joint possibility.

\looseness=-1 \para{\bf Goals and Challenges.} In this run-time reference stage, {\bf \adv's goal} is to \erev{use $\mathcal{C}$} to identify the face in a query image. {\bf \pr's goal} is to alter their query image so \erev{$\mathcal{C}$ cannot match it} to their corresponding feature vector in $\mathbf{D}$.  We assume \adv's reference database $\mathbf{D}$ contains accurate feature vectors of \pr. 

There are two key challenges for stage $\crc{5}$-based AFR proposals. The first is achieving successful evasion without significant image distortion. Additionally, proposals must overcome defenses deployed by \adv~to protect $\mathcal{F}$ and/or $\mathcal{C}$.

\vspace{-0.1cm}
\subsection{\bf Current Solutions}
\vspace{-0.1cm}

Adversarial perturbations have been the dominant
method for evading DNNs and consequently are relevant
for evading FR. Due to the extremely high number of these
techniques, we restrict our discussion to proposals explicitly
designed to evade FR systems at run-time.  We organize these proposals by their operational context: physical and digital. 

\looseness=-1 \para{Physical evasion techniques.}  The first group of proposals
injects adversarial perturbations into face images by having \pr~wear
them as physical objects.  While these methods echo those described in
\S\ref{sec:preprocess},  they focus on thwarting recognition/classification rather than face detection.  Earlier proposals~\cite{accessorize,
  feng2013facilitating} use adversarial makeup and eyeglasses to cause
incorrect classification by \erev{$\mathcal{C}$}. More recent proposals consider
two other directions, either using larger but input-independent adversarial patches to boost the effectiveness of evasion~\cite{advhat}, or making the perturbation digitally controllable and/or much less perceivable by human eyes by projecting
visible/infrared light onto user faces~\cite{vla, invisiblelight,
  adversariallight}.   

\para{Digital evasion techniques.}  Here \pr~digitally modifies 
their unlabeled (online) face images to prevent them from being accurately classified by \erev{$\mathcal{C}$}.  Most proposals in this category apply traditional adversarial perturbation generation techniques to create minimally visible perturbations that cause \adv's feature extractor to produce misleading feature vectors.  Their generation process varies depending on \erev{assumptions of $\mathcal{C}$'s behavior}:  a shallow classification layer vs. nearest neighbor based matching~\cite{makeupevasion, fastperturb, efficientblackbox, singh2021brightness}.

More recent proposals are designed to be more robust to
real-world FR systems (i.e. joint optimization on multiple feature
extractors, etc.)~\cite{faceoff,
  preventing,yang2021towards}. Another recent 
proposal~\cite{advfaces} uses a GAN to generate adversarial perturbations rather than using optimization techniques.

\rev{
\vspace{-0.1cm}
\subsection{\bf Discussion of Stage $\crc{5}$ Solutions}
\vspace{-0.1cm}

\para{Privacy/Utility Trade-offs.} For physical evasion techniques, a significant usability challenge comes from the possibility of real-time recognition. In order to ensure physical evasion tools are effective, a user must wear them in all circumstances where cameras might be present. As with Stage $\crc{4}$, there also exists here a trade-off between perturbation size and evasion success for digital evasion techniques. Reducing the amount of perturbation needed to evade recognition remains an active area of research. 

\looseness=-1 \para{Summary of Progress.} So far, existing works have focused on addressing the first key challenge, and have evasion success with minimally visible perturbations on query images. Addressing the second challenge, or understanding how AFR tools interact with existing defenses against evasion attacks, remains an open area of research. Defenses against evasion attacks like the ones listed above are being released regularly (e.g.~\cite{meng2017magnet,samangouei2018defense, dhillon2018stochastic, madry2017towards}), only to be broken by new attacks (e.g.~\cite{carlini2017magnet, athalye2018obfuscated, tramer2020adaptive}). No defenses have yet been explicitly proposed for these attacks, but the general trend suggests this may be possible. 

}

\begin{table}
\centering
\arrayrulecolor{black}

\resizebox{0.5\textwidth}{!}{
\begin{tabular}{cccccc}
  \toprule
\multicolumn{1}{c}{\multirow{3}{*}{\begin{tabular}[c]{@{}c@{}}\\\textbf{ Stage}\\\textbf{Targeted}\end{tabular}}} & \multicolumn{5}{c}{\textbf{AFR Property}} \\ \cmidrule{2-6}
\multicolumn{1}{c}{} & \multicolumn{1}{c}{\begin{tabular}[c]{@{}c@{}}\textit{Long-term~}\\\textit{Robustness}\end{tabular}} & \multicolumn{1}{c}{\begin{tabular}[c]{@{}c@{}}\textit{Broad}\\\textit{Coverage}\end{tabular}} & \multicolumn{1}{c}{\begin{tabular}[c]{@{}c@{}}\textit{No 3rd Party}\\\textit{Assistance}\end{tabular}} & \multicolumn{1}{c}{\begin{tabular}[c]{@{}c@{}}\textit{Disruption}\\\textit{to \pr}\end{tabular}} & \multicolumn{1}{c}{\begin{tabular}[c]{@{}c@{}}\textit{Disruption~}\\\textit{to others}\end{tabular}} \\ 
\midrule
$\crc{1}$ & \gcirc & \ecirc & \ecirc & \gcirc & \fcirc \\ 
%\hline
$\crc{2}$ & \gcirc & \ecirc & \fcirc  & \gcirc & \fcirc \\ 
%\hline
$\crc{3}$ & \ecirc & \ecirc & \ecirc  & \gcirc & \ecirc \\ 
%\hline
$\crc{4}$ & \gcirc & \gcirc & \fcirc & \gcirc & \ecirc \\ 
%\hline
$\crc{5}$ & \ecirc & \fcirc & \fcirc & \gcirc  & \ecirc \\
%\arrayrulecolor{black}\hline
\bottomrule
\end{tabular}
}
\vspace{0.1in}
\caption{Evaluating AFR tools using five properties, where the tools
  are grouped by the FR stage they target.
  }
\label{table:property}
\end{table}

%\vspace{0.08in}
\section{\bf Goals and Tradeoffs in AFR Design}
\label{sec:tradeoffs}

\looseness=-1  In our discussion of current AFR tools, we consider the design space of AFR
tools through the lens of specific FR stages they disrupt. To date, all
existing AFR proposals have focused their design around
disrupting a single stage in this framework. Assuming an AFR tool must
disrupt some portion of the FR pipeline to be effective, we can map out and
explore the design space of AFR tools using this framework.

For researchers and practitioners in the AFR community, perhaps the most
critical question is: {\em ``what are the benefits and limitations of AFR tools that target
each specific framework stage?''} Or, an alternative form of the
question might be: {\em ``Given a set of prioritized properties for an AFR system, can I
find the best stage(s) to disrupt in order to achieve them?''}

\looseness=-1 We attempt to answer these questions here, by first identifying a set of high
level properties that AFR tools can potentially optimize for, then for each
property, discussing how targeting a given stage affects an AFR tool's
ability to achieve it. Ultimately, we hope to provide a high level
roadmap that can guide the design of AFR tools optimizing for specific properties
in mind. While we consider each stage in isolation, it might be
possible for an AFR tool to target multiple stages, gaining a
combination of benefits (and limitations).

\vspace{-0.2cm}
\subsection{\bf Five AFR Design Properties} % to Consider for AFR Design}
\vspace{-0.2cm}

When considering design properties of AFR tools, we assume that efficacy 
is a given. Our list of 5 properties target additional considerations
beyond basic efficacy, and include desirable properties for efficacy
(\#1 and \#2) and for minimizing dependencies and cost (\#3,
\#4, \#5):

\begin{packed_enumerate}
\item {\em Long-term robustness} against evolving FR systems
\item {\em Broad protection coverage}, efficacy even for users with
  unprotected face images online
\item {\em No reliance on 3rd parties}, does strong protection
  require assistance from service providers or others?
\item  {\em Minimal friction for user \pr}, minimizing cost for $P$ to
  deploy the AFR tool on a consistent basis
\item {\em Minimal impact on other users}, minimizing potential risks to
  non-users of the AFR tool
\end{packed_enumerate}

\vspace{-0.2cm}
\subsection{\bf Implications of Properties for AFR Design}
\vspace{-0.2cm}

Next, we discuss the above properties in turn and consider how easily each
property can be achieved by AFR tools that target different operational
stages in our framework. For each combination of property and target stage, we ``quantify'' how easily
the desirable property can be achieved by an AFR tool designed to disrupt that stage.
$\fcirc$ means that the property has already been achieved by current AFR
proposals targeting this stage; $\gcirc$ means that the property seems ``promising''
and has good potential to be achieved by AFR designs targeting this stage;
and $\ecirc$ indicates significant progress may be required to achieve this
property by targeting this stage, and the likelihood of success is unknown.  Table~\ref{table:property} provides an overview of our conclusions. For easy
notation, we will use {\bf AFR$\crc{k}$} to refer to the group of AFR
proposals that target FR stage {\bf $\crc{k}$}.

\noindent\rule{\linewidth}{0.4pt}
\noindent {\bf Property 1: Long-term robustness.}
An effective AFR tool should provide strong and lasting protection against
unwanted facial recognition. That is, it should protect a user \pr~from
unwanted FR both initially and as FR evolves.

\para{$\fcirc$: None} While this principle is the main goal of AFR, none of existing AFR
tools (targeting any stage) is able to achieve this property. 
No current system provides strong protection against ever-evolving FR systems. 

\para{$\gcirc$: AFR$\crc{1}$, AFR$\crc{2}$, AFR$\crc{4}$} Conceptually,
\pr~can achieve long-term robustness by consistently undermining the face
data pipeline of \adv. AFR$\crc{1}$ and AFR$\crc{2}$ can both prevent any
face image of \pr~to be included into \adv's pipeline.  AFR$\crc{4}$ can
corrupt \adv's understanding of any face images in the reference
database. While promising, existing AFR tools fail to {\em consistently}
prevent the inclusion of or corrupt {\em all} \pr's images from both online
and physical sources.

\para{$\ecirc$: AFR$\crc{3}$, AFR$\crc{5}$} It remains unclear if these two
groups of AFR tools can provide long-term robustness. AFR$\crc{3}$ could be
overcome over time as \adv~switches to newer and different feature
extractors. AFR$\crc{5}$ offers only one-time protection, and does not
address the scenario where query images get added to the reference database.

\noindent\rule{\linewidth}{0.4pt}
\noindent {\bf Property 2: Broad protection coverage.}
Many of us already have an online presence, \eg face photos posted years ago without 
AFR protection. An effective AFR proposal would ideally provide protection
under the challenging but realistic scenario where \pr~already has unprotected face 
images online. 

\para{$\fcirc$: AFR$\crc{5}$ } AFR tools that rely on run-time evasion are not impacted by 
the existence of unprotected images online. 

\looseness=-1 \para{$\gcirc$: AFR$\crc{4}$ } The presence of unprotected images complicates the protection
of AFR$\crc{4}$ since \adv~has some ground truth information about
\pr's facial features. However, the addition of protected images to the reference database can slowly disguise \pr's true features, and thus achieve protection. Moreover, several AFR 
tools~\cite{fawkes,foggy} proposed a ``group cloaking'' idea where
multiple users coordinate together to
achieve better protection for those having an existing online presence. 

\para{$\ecirc$: AFR$\crc{1}$, AFR$\crc{2}$, AFR$\crc{3}$ } These three groups of AFR tools focus
on disrupting the (training) data pipeline of FR. As a result, they cannot protect \pr~against 
\adv~who has already obtained and processed unprotected images of \pr.

\noindent\rule{\linewidth}{0.4pt}
\noindent {\bf Property 3: No reliance on 3rd party to operate.}
Ideally, an AFR tool can be operated by a user \pr~alone and achieve strong protection without assistance or participation third-party, either a central content provider like Facebook or a friendly user willing to help \pr.  This is an abstract measure of the entity-level complexity required to operate the tool.  Achieving this property has the added benefit of limiting exposure of potentially sensitive user data to a 3rd party.%, {\em i.e.\/} the AFR is also privacy-preserving.

\para{$\fcirc$: AFR$\crc{2}$, AFR$\crc{4}$, AFR$\crc{5}$} AFR tools in
these three groups all rely on adding 
certain perturbations on face images, which \pr~can do without outside assistance.

\para{$\ecirc$: AFR$\crc{1}$, AFR$\crc{3}$}  For those AFR$\crc{1}$
seeking to prevent online data scraping,  they rely on the assistance
of image sharing platforms.  Similarly, disrupting the training $\mathcal{F}$ requires coordinated effort across many users,
since \pr~alone contributes relatively few images to the training data.

\noindent\rule{\linewidth}{0.4pt}
\noindent {\bf Property 4: Minimal friction for \pr.} 
This usability-related property measures what \pr~needs to sacrifice
in order to consistently apply the AFR tool.  This property is motivated by the well-known findings that users prefer and are
more likely to use protection solutions that introduce minimal friction to their daily life~\cite{coopamootoo2020usage,wright2002privacy}.

\para{$\gcirc$: AFR$\crc{1}$, AFR$\crc{2}$, AFR$\crc{3}$, AFR$\crc{4}$, AFR$\crc{5}$ }
So far, existing AFR tools all introduce some level of
``disruption'' to \pr, whether by adding visual noise,
perturbations or transformations to \pr's online photos that distorts them, requiring \pr~to always wear odd 
makeup/clothes/accessories, or necessitating more powerful computing
hardware/services to implement the AFR tool against continually evolving \adv.  More research efforts are needed to limit the amount/type of disruption to users.

\noindent\rule{\linewidth}{0.4pt}
\noindent {\bf Property 5: Minimal impact on other users.}
This final property examines how the outcome of \pr's AFP protection would affect other users.  Intuitively, \pr~can protect themselves by forcing \adv~to fail (give a null or uninformative result), or by intentionally tricking \adv~to recognize them as another person \pr'.  Depending on the context, the latter may negatively affect \pr', producing potential social risks (see \S\ref{subsec:social} for detailed discussions on social challenges of AFR).

\para{$\fcirc$: AFR$\crc{1}$, AFR$\crc{2}$ } These AFR tools disrupt the data pipeline of \adv, and thus, have no impact on other users. 

%\para{$\gcirc$: None }

\para{$\ecirc$: AFR$\crc{3}$, AFR$\crc{4}$, AFR$\crc{5}$ } These three groups
of AFR tools seek to intentionally misclassify \pr's face to another user,
and as a result, could potentially impact other users included in
\adv's reference database.

\section{\bf Challenges for AFR Tools}
\label{sec:future}
\vspace{-0.1cm}

In this section, we describe what we see as the major technical and broader
social/ethical challenges facing future AFR development. Each challenge spans
multiple properties and stages laid out in this paper. For each challenge, we
provide context for why the challenge exists and, where possible, suggest
ways to address it. Like \S\ref{sec:tradeoffs}, the challenges described here
represent our best efforts to understand and systematize the AFR space. They
are not exhaustive, and are meant as signposts rather than a comprehensive
roadmap. % for future research.

\vspace{-0.2cm}
\subsection{\bf Technical Challenges}
\label{subsec:tech}
\vspace{-0.2cm}

%\noindent\rule{\linewidth}{0.4pt}
\para{TC 1: Reliance on AML-based tools.} The majority of AFR proposals, especially those targeting stages
  {\bf $\crc{2}-\crc{5}$},  employ techniques from adversarial machine learning (AML), which have several key limitations. First, while AML tools have exhibit high performance, they have not provided provable guarantees of protection. %For example, te success rate of adversarial perturbations may drop significantly when \pr's knowledge of \adv~is imperfect~\cite{demontis2019adversarial}.  
  Second (and related), AML-based protections can be defeated by adaptive FR systems. For example, \adv~could adversarially train the feature extractor $\mathcal{F}$~\cite{oriole, savior} to
  be more robust against adversarial examples,  thus defeating AFR
  tools against stages $\crc{3}$ or $\crc{4}$. \adv~could also remove adversarial perturbations from face images before processing them or adding them  to the reference database~\cite{faceguard}, circumventing AFR tools that target stages $\crc{2}$ or $\crc{5}$.

{\em Potential Directions.}  More advanced perturbation generation methods may help increase short-term efficacy of AML-based AFR
tools. However, the lack of provable, ongoing
protection is a much tougher barrier to overcome. 
In order to provide reliable, ongoing protection, developers of AFR tools can consider two possible paths: (i) integrate provable guarantees into the perturbation generation process, or (ii) consider alternative techniques that provide guaranteed protection.  For (ii), there are two potential directions. The first is focus on attacking stage $\crc{1}$, where defeating FR does not require evading or poisoning a DNN (e.g. non-AML AFR tools).  The second is to switch from ``misleading'' \adv~with ``minor'' image modifications to completely disabling $\mathcal{F}$ and/or $\mathcal{C}$. \rev{Methods in this direction could focus on physical world attacks that exploit camera properties, like the rolling shutter effect~\cite{liang2008analysis, sayles2021invisible}; rely on larger image disruptions like shadows~\cite{zhong2022shadows}; or employ tools like the FR-disabling lasers used in Hong Kong in 2020~\cite{hongkong}.}

%\noindent\rule{\linewidth}{0.4pt}
\looseness=-1 \para{TC 2: Existence of online footprints.} Some AFR proposals (especially those targeting stage $\crc{4}$) implicitly or
explicitly assume that users can start ``from scratch'' to protect their
online persona. In practice, most Internet users today already have face
images online, posted by themselves or others, and at least some of those
images are already captured by FR databases. Over 1.8 billion photos are
uploaded to online platforms daily~\cite{forbes_photos}, making it likely that one or more unmodified photos of a user \pr~will likely end up online, with or without \pr's knowledge. Given the widespread use of web scraping to
collect FR reference images~\cite{pimeyes, hill_clearview}, it is likely that one of these photos is already in a reference database.

\looseness=-1  {\em Potential Directions.}  This stark reality has two implications for AFR research. {\em First}, AFR tools should be evaluated under the
practical scenarios where the FR system has access to both protected and
unprotected online photos of \pr.  While several AFR tools have provided such
measurements (\eg~\cite{fawkes, foggy}), many others have not. {\em
  Second}, we believe that AFR tools managed by online platforms will offer better protection of online footprints against FR systems than those executed by individual users. These platforms can protect photos of an individual
posted by them or others, and are overall better positioned to deploy more powerful protection mechanisms.

For example, online platforms could employ the group cloaking techniques proposed in Fawkes~\cite{fawkes} or FoggySight~\cite{foggy} to corrupt reference databases 
 composed of images from their sites. After images are scraped,
 online platforms could use provenance-tracking to re-identify stolen images,
\eg in the training dataset of a feature extractor, and enable
exposure/prosecution of photo thieves~\cite{salem2018ml,shokri2017membership,
  radioactive}. All these methods ought to be accompanied by enhanced
anti-scraping techniques to prevent large-scale scraping of face
images, \ie stricter rate limiting, access permissions, and scraping
detection heuristics, to make it safer for individuals to have online footprints.

\para{TC 3: Privacy/utility/usability tradeoffs of AFR systems.} The above paragraph raises an additional 
technical challenge of AFR design: balancing privacy, utility, and usability. There is a spectrum of ways 
to balance these. On one end are 3rd-party-adminstered AFR tools (c.f. stage~$\crc{1}$), which have the 
high usability and utility but intrude on privacy to allow 3rd party data processing. On the other end are 
high-overhead tools like fully homomorphic encryption, which have provable privacy but limited 
utility/usability.

\looseness=-1  Where AFR tools should or can exist along this spectrum remains an open question for two reasons. First, 
we lack a deep understanding of how AFR users would prioritize these tradeoffs in practice. Prior 
studies show that users prefer privacy tools with minimal 
overhead~\cite{coopamootoo2020usage,wright2002privacy}, but only one study has explored how or if these 
preferences change in the AFR setting~\cite{faceoff}. Second, while many AFR tools have been proposed in recent years, the 
space of possible AFR designs remains sparsely populated. Consequently, it is worth considering whether 
this tradeoff is indeed fundamental, or if future AFR designs may evolve to accommodate 
all three.

\para{TC 4: Face images don't change.} A related, but distinct, challenge to TC 2 faced by AFR systems is the
permanence of face data.  For better or for worse, most people have the same
face their whole adult life.
As our faces age, they remain recognizable as uniquely ``us'' to most
humans and FR systems~\cite{ling2009face}. The slow rate at which faces change is a
major challenge for AFR tools. To be long-term effective, these tools
must conceal the same piece of static data (a face) from numerous
adversaries over many years.

Once \adv~obtains \pr's protected face photo, they can try as many times as they want to break the
protection~\cite{savior}. If \adv~ever succeeds, either in 1 month or 1 year, they ``win'' and
\pr~loses, because modern FR systems only need one clean picture in the reference
database to identify a person~\cite{arcface}. For example,
Clearview.ai identified a person based on a single reference image in which
the person's reflection appeared faintly in a mirror~\cite{hill_clearview}.
Clearly, the issue of face data permanence poses a significant
challenge for AFR tool development.

\para{TC 5:  Lack of transparency of FR systems.} The lack of transparency in how proprietary FR systems work hampers AFR
tool development and testing. Without access to proprietary FR systems, AFR
researchers must do their best to glean a generic understanding of how FR
systems work from public documents and academic papers, e.g.~\cite{facenet,
  gao_2020}. While this may be sufficient to develop AFR tools that work well
in the lab, researchers cannot perform comprehensive efficacy tests against proprietary systems.

Furthermore, AFR tool developers have no knowledge of how or if FR systems
are adapting to evade AFR systems. The 2020 global FR market was
valued at 3.86 billion US dollars~\cite{global_money}, so FR stakeholders have
ample resources to evolve as new AFR systems
emerge. Even passive improvements to FR systems, such as new
training methods or architectures, can overcome AFR protection and compromise
user privacy~\cite{savior}. Altogether, this lack of transparency means
that AFR tools face an upward battle in the fight against unwanted FR.

\vspace{-0.2cm}
\subsection{\bf Broader Social and Ethical Considerations}
\label{subsec:social}
\vspace{-0.2cm}

\looseness=-1 In addition to technical challenges, AFR tools
face broader social and ethical considerations. These stem from a variety of
factors, including a lack of regulation, benefits of FR for the
public good, and demographic disparities in FR. %$ systems.

\para{SEC 1: Unregulated, ubiquitous FR.} Today, FR systems are generally unregulated and easy to deploy.
Practically anyone  with a powerful laptop and access to an image
dataset could create a FR system.  This democratization of FR has allowed 3rd party FR systems like Clearview.ai, which rely on unauthorized data
 use~\cite{clearview_explained}, to flourish.  As a result, it is
 extremely difficult (if not impossible) for individuals to know
 when and where FR systems are deployed, as well as their capabilities.

This laissez-faire climate creates significant ambiguity as to when AFR tools can/should be deployed.
For example, around the world, photos taken for official government
purposes ({\em e.g.\/} drivers' license and passport photos) are used as
reference images in government FR systems aiding law enforcement
officers,  border control agents, among others~\cite{perpetual, israel_fr, cbp_facialrec,
  mozur2019one}. Government-sponsored FR may be {\em unwanted} but
is not (necessarily) {\em unauthorized} under the status quo, and the legality of using AFR tools in this setting is ambiguous. To augment the confusion, systems like Clearview are used by law
enforcement~\cite{hill_clearview}, further blurring the concept of {\em
  unauthorized} vs {\em unwanted} FR. As FR and AFR use increases, a clash over this issue seems almost inevitable. 

\para{SEC 2: FR used for social good.} Both privacy-sensitive citizens and criminals can use AFR tools. Law
enforcement's use of facial recognition can benefit society in multiple ways,
such as tracking and locating wanted criminals or lost
children~\cite{found_child, faceslist}. Consequently, AFR tools applied by
bad actors could ultimately harm the public good. The debate between privacy and national
security plays out in numerous other tech domains, such as end-to-end
encryption~\cite{doj_e2e}. Legitimate claims can be made by both sides. AFR
researchers must be mindful of this tension and the
potential consequences of their work.

%\noindent\rule{\linewidth}{0.4pt}
\para{SEC 3: Harm caused by AFR misidentification.} One ethical tension not yet explored in current literature is the social effect of misidentifications
caused by AFR tools. For example, if $U$ uses an AFR tool and is
misidentified by \adv~as $P$, what outcome might this
have for $P$? If $U$ is engaging in illegal activity but $P$ is
arrested instead, the AFR tool could cause serious harm, both to $P$
and to $U$'s victim(s). The well-known bias of FR systems heightens this tension. Police departments routinely make rushed identification decisions from partial FR matches~\cite{garbagein}. Furthermore, facial recognition systems misidentify people of color at higher
rates~\cite{gendershades, dooley2021robustness}. Recent work has found
that AFR tools exhibit these same biases~\cite{rosenberg2021fairness,
  qin2021bias}. The social impact of AFR tools requires urgent study.

\vspace{-0.1cm}
\section{\bf Discussion}
\vspace{-0.2cm}
\label{sec:discussion}

\para{Related Surveys.} Two surveys~\cite{meden2021privacy, padilla2015visual}, alluded to in 
\S\ref{sec:intro}, address topics similar to our work. Here, we provide an in-depth comparison to these and 
emphasize our unique contributions. ~\cite{meden2021privacy} focuses on how a {\em service provider} can build 
a privacy-preserving FR service (see it's \S II.D), while our work considers how {\em individual users} can use 
a privacy tool to defend themselves against {\em intrusive} FR services.  Different 
from~\cite{meden2021privacy}, our work provides a holistic view of what individual users can do to disrupt FR 
and a detailed discussion of challenges facing user-centric AFR solutions.

\looseness=-1 \cite{padilla2015visual} also discusses methods a service provider could use to preserve privacy in a video surveillance setting. 
Furthermore, since ~\cite{padilla2015visual} focuses on video surveillance, rather than image-based FR systems, it explores privacy preservation techniques for video-specific traits like gait, height, and clothing, which dilutes FR-specific content. ~\cite{padilla2015visual} does address face de-identification in its \S3.4.3, but focuses solutions for providers who wish to deanonymize {\em all} faces in their system (e.g. by averaging or blurring them), unlike our focus on protecting individuals from unwanted surveillance.

\para{Concluding Thoughts.} As facial recognition (FR) continues to grow in scale and ubiquity, we expect
the demand for anti-facial recognition tools to continue to rise.  There is an urgent need to
think longitudinally about AFR tools, analyzing both their limits and their
potential. Our paper aims to fill this gap by providing both a framework for
discussing AFR proposals and an assessment of the current state of AFR
research.

Current AFR tools possess some, but not all, of the traits
needed to successfully defeat unwanted FR in the real world. Many
proposals leverage adversarial perturbations to evade FR models, either in
the preprocessing $\crc{2}$ or classification $\crc{5}$ stages. These are often effective in the short-term, but lack long-term guarantees, and cannot fundamentally change FR system behavior in the future. Future AFR proposals may benefit from more exploration of designs that target stages $\crc{1}$ and $\crc{4}$, which could provide
wider-reaching protection.

\para{Acknowledgements.}  We thank our anonymous reviewers and shepherd for their insightful feedback. This work is supported in part by NSF grants CNS1949650, CNS-1923778, CNS-1705042, by C3.ai DTI, and
by the DARPA GARD program.
Emily Wenger is supported by a GFSD Fellowship and
a Harvey Fellowship. Both Emily Wenger and Shawn Shan
are also supported by Neubauer Fellowships at the University of Chicago. Any opinions, findings, and conclusions or recommendations expressed in this material are those of the authors and do not necessarily reflect the views of any funding agencies.

\newpage

\small
\bibliographystyle{IEEEtran}
\bibliography{references}

\appendix
\section{Appendix}

Tables~\ref{tab:government_use} and~\ref{tab:labeled_data} provide more context for the FR deployment and data collection content of \S2.3. Table~\ref{tab:government_use} lists known uses cases of FR systems by governments around the world. Table~\ref{tab:labeled_data} lists known sources of labelled data for well-known large-scale FR systems, both public and private. 
%\vspace{-0.5cm}

\begin{table}[h]
    \centering
    %\vspace{-6cm}
    \resizebox{0.5\textwidth}{!}{
      \begin{tabular}{ cll } 
        \toprule
        \multicolumn{1}{c}{\textbf{Location}} &
                                               \multicolumn{1}{c}{\textbf{Use
                                               Cases Reported}} &
                                                                  \multicolumn{1}{c}{\textbf{Countries/Companies
                                                                  }} \\ 
        \midrule
        \multirow{6}{*}{\begin{tabular}[c]{@{}l@{}}\\ \\ \\ \\ Public \\ spaces\end{tabular}} %& Social credit monitoring & China~\cite{mozur2019one} \\ \cline{2-3}
        & On-street surveillance & \begin{tabular}[c]{@{}l@{}}
                                 Bahrain~\cite{saudi_bahrain_rec},
                                 China~\cite{uighur_surveill},
                                 England~\cite{bigbrother}, \\   
                                 France~\cite{france_fr}, 
                                 Kenya~\cite{kenya_police},   
                                 Myanmar~\cite{bel_gr_my}, \\
                                 Russia~\cite{abc_russia}, 
                                 UAE~\cite{saudi_bahrain_rec}, 
                                 UK~\cite{london_police_fr}, \\
                                 US~\cite{facesurveillance}, 
                                 Zimbabwe~\cite{jie2018china}\end{tabular}
        \\ \cmidrule{2-3}
        & Criminal suspect identification & \begin{tabular}[c]{@{}l@{}}
                                              Argentina~\cite{argentina_fr},
                                              Belarus~\cite{bel_gr_my}, \\
                                              Brazil~\cite{hurtling}, 
                                              China~\cite{china_ai}, \\
                                              Greece~\cite{bel_gr_my}, 
                                              Malaysia~\cite{malaysia_fr},
                                              US~\cite{perpetual}
                                            \end{tabular} \\ \cmidrule{2-3} 
         & School monitoring & \begin{tabular}[c]{@{}l@{}} Brazil~\cite{school_fr},
                               China~\cite{andrejevic2020facial},
                               India~\cite{india_monitoring}, \\
                               Russia~\cite{school_fr},
                               US~\cite{school_fr}  \end{tabular} \\ \cmidrule{2-3}
        & Border security &  Israel~\cite{israel_fr},
                             Pakistan~\cite{pakistan_border},
                             US~\cite{cbp_facialrec} \\ \cmidrule{2-3}
                                           & COVID lockdown enforcement
                                                                &\begin{tabular}[c]{@{}l@{}}
                                                                  China~\cite{china_covid},
                                                                  India~\cite{covid_asia}, \\
                                                                  South Korea~\cite{covid_asia},
                                       Russia~\cite{abc_russia} \end{tabular} \\
        \midrule 
        \multirow{4}{*}{\begin{tabular}[c]{@{}c@{}} \\  Privatized \\ spaces\end{tabular}} & Catching shoplifters& Apple,
                                                            Macy's,
                                                            Lowe's~\cite{apple,insider}
        \\ \cmidrule{2-3}
     & Securing facility access & Alibaba~\cite{alibaba},
                                       Intel~\cite{intel} \\ \cmidrule{2-3} % , \\ Madison
        & Tracking driver behavior & Hyundai~\cite{hyundai},
                                     Subaru~\cite{subaru}
                                     % Square Gardens~\cite{madison}                                   
        \\ \cmidrule{2-3}
        & Airline passenger check-in & JetBlue~\cite{jetblue}, Delta~\cite{delta} \\ \bottomrule
                                  %\
        %                              % Employee happiness tracking & McDonalds Japan~\cite{mcdonalds} \\ \hline
        %                              % Meal order prediction & KFC China~\cite{kfc} \\ \hline
                                     % Virtual makeup try-on & Mac Cosmetics~\cite{mac} \\ \hline
        \end{tabular}
      }
      \vspace{0.1cm}
      \caption{Example use cases of facial recognition.}
        %\vspace{-0.15in}
      \label{tab:government_use}
    %\end{table}

%\begin{table}[t]
\centering
\begin{tabular}{ll} 
\toprule
\multicolumn{1}{l}{\textbf{Operator of FR system}} &
                                                \multicolumn{1}{l}{\textbf{
                                                                Source
                                                                of reference images}} \\
    \midrule
Clearview.ai & Social media photos~\cite{hill_clearview} \\
PimEyes & (Public) online photos~\cite{pimeyes} \\
FBI F.A.C.E.S. & State drivers' license photos~\cite{perpetual}
    \\
US Customs and Border Patrol & Passport photos~\cite{cbp_facialrec}  \\
Skynet (China) & National ID photos~\cite{chinafile_shawan, mozur2019one} \\
    \bottomrule
\end{tabular}
\vspace{0.03in}
\caption{Reported reference image sources} 
\label{tab:labeled_data}
\end{table}

\end{document}